\documentclass[twocolumn,showpacs,amsmath,amssymb,prd]{revtex4-1}

\usepackage{amssymb,amsmath,latexsym,mathrsfs}
\usepackage{graphicx}
\usepackage{epsfig}
\usepackage{soul}

\newcommand{\beq}{\begin{equation}}
\newcommand{\eq}{\end{equation}}
\newcommand{\bear}{\begin{eqnarray}}
\newcommand{\ear}{\end{eqnarray}}

\newcommand{\od}{\Omega_{\rm de}}
\newcommand{\omegae}{\Omega_{\rm e}}
\newcommand{\ome}{\Omega_{\rm e}}

\newcommand{\aeq}{a_{\rm {eq}}}
\newcommand{\dalfa}{$\Delta\alpha/\alpha$ }

\begin{document}


\title{Dark Energy coupling with electromagnetism as seen from future low-medium redshift probes }

\author{E. Calabrese$^1$, M. Martinelli$^{2}$, S. Pandolfi$^3$,\\ V. F. Cardone$^4$, C. J. A. P. Martins$^{5}$,  S. Spiro$^6$, P. E. Vielzeuf $^{5,7}$}

\affiliation{$^1$Sub-department of Astrophysics, University of Oxford, Keble Road, Oxford OX1 3RH, UK}
\affiliation{$^2$SISSA, Via Bonomea 265, Trieste, 34136, Italy}
\affiliation{$^3$Dark  Cosmology  Centre,  Niels  Bohr  Institute,  University  of  Copenhagen,  Juliane  Maries  Vej  30,  DK-2100  Copenhagen, Denmark.}
\affiliation{$^4$I.N.A.F.\,-\,Osservatorio Astronomico di Roma, via Frascati 33, 00040 - Monte Porzio Catone (Roma), Italy }
\affiliation{$^5$Centro de Astrof\'{\i}sica da Universidade do Porto, Rua das Estrelas, 4150-762 Porto, Portugal}
\affiliation{$^6$I.N.A.F.\,-\,Osservatorio Astronomico di Padova, Vicolo dell'Osservatorio 5, 35122 Padova, Italy}
\affiliation{$^7$Faculdade de Ci\^encias, Universidade do Porto, Rua do Campo Alegre 687, 4169-007 Porto, Portugal}

\begin{abstract}
Beyond the standard cosmological model the late-time accelerated expansion of the universe can be reproduced by the introduction of an additional dynamical scalar field. In this case, the field is expected to be naturally coupled to the rest of the theory's fields, unless a (still unknown) symmetry suppresses this coupling. Therefore, this would possibly lead to some observational consequences, such as space-time variations of nature's fundamental constants. In this paper we investigate the coupling between a dynamical Dark Energy model and the electromagnetic field, and the corresponding  evolution of  the fine structure constant ($\alpha$) with respect to the standard local value $\alpha_0$. In particular, we derive joint constraints on two dynamical Dark Energy model parametrizations (the Chevallier-Polarski-Linder and Early Dark Energy model) and on the coupling with electromagnetism $\zeta$, forecasting future low-medium redshift observations. We combine supernovae and weak lensing measurements from the Euclid experiment with high-resolution spectroscopy measurements of fundamental couplings and the redshift drift from the European Extremely Large Telescope, highlighting the contribution of each probe. Moreover, we also consider the case where the field driving the $\alpha$ evolution is not the one responsible for cosmic acceleration and investigate how future observations can constrain this scenario.
\end{abstract}
\date{\today}

\maketitle
\section{Introduction}
\label{sec:intro}

Since the discovery of cosmic acceleration from measurements of luminosity distances of type Ia Supernovae (SN) in 1998~\cite{Perlmutter:1998np,Riess:1998cb} and its confirmation by several other independent cosmological data, the nature of the component driving this acceleration, the so-called Dark Energy (DE hereafter), has been deeply debated. In the standard cosmological model, the $\Lambda$ Cold Dark Matter ($\Lambda$CDM), the acceleration is produced by the cosmological constant $\Lambda$. This model is consistent with the majority of the observational data, but the known theoretical problems of the cosmological constant led cosmologists to formulate several other alternative models able, from one side, to relieve the aforementioned theoretical issues and, on the other side,  to explain observations.

Alternative models for the DE, such as quintessence, are called (models of) dynamical dark energy and, even if not favoured, they are currently not excluded by observations \cite{Ade:2013lta,Said:2013jxa}. Several of these alternative models are characterized by the existence of an additional scalar field which drives the accelerated expansion of the universe. If this is the case, it is expected that this additional component is coupled to the rest of the theory's fields.

In this paper we study the coupling of dynamical DE models with the electromagnetic field: indeed, the presence of this coupling would lead to a space-time variation of the fine-structure constant $\alpha$ \cite{Carroll}. This, in turn, would generate distinctive signatures in cosmological data, such as the Cosmic Microwave Background (CMB) (see e.g.~\cite{Kaplinghat,Fisher,Menegoni:2009rg,Calabrese:2011nf}), but also in low and medium redshift probes, for example in the peak of luminosity in SN or in the metal absorption lines of distant quasars (QSO).

The present work aims to extend and to complete the analysis done in \cite{Calabrese:2011nf}, where constraints on the coupling of a time-varying fine structure constant in the presence of Early Dark Energy were obtained with CMB data. In this paper we focus on low-medium redshift observables, forecasting  SN and QSO data, Weak Lensing shear power spectrum measurements (WL), and redshift-drift (RD) data. 
The relevance of this combination of probes is the coverage of a wide redshift range ($0<z\lesssim 5$) which is a very powerful way to discriminate between a cosmological constant and a dynamical DE model, as it makes possible to investigate the onset of DE. In other words, given the possibility of a dynamical field that is moving very slowly (in appropriate units), searching for deviations from a cosmological constant is optimally done by maximizing the lever arm of probed redshifts.

In the present work we assume only a time-varying fine structure constant, neglecting spatial variation. Recent analyses of CMB data \cite{spatialcmb} have shown no evidence of a spatial variation; there is instead some evidence of a spatial variation from lower redshift QSO measurements \cite{Webb}, and attempts are being made to independently confirm it \cite{LP1,LP2}. For the moment we note that our method could in principle be extended to the more complex models needed to account for such spatial variations.

We consider two different classes of time-varying $\alpha$ models. In the first class the scalar field causing the $\alpha$ variation is also responsible for the accelerated expansion of the universe, and therefore observational tests of the evolution of $\alpha$ directly contribute to constrain dark energy scenarios \cite{Amendola:2011qp}. In the second class the additional degree of freedom which causes the $\alpha$ variation is not (or at most is only partially) the source of the DE component. This second class is important for two reasons. Firstly, although consistency tests are available, erroneous dark energy properties could be inferred if the $\alpha$ evolution is ascribed to DE instead of this ``external'' degree of freedom; this scenario has been discussed in \cite{Vielzeuf:2012zd}. Moreover, there may be a bias induced on the cosmological parameters estimation due to a wrong assumption on the underlying cosmological model, i.e. selecting a dataset with a non-zero variation of $\alpha$, but assuming no variation in the analysis. We investigate this possibility here. Should such a bias be non-negligible and found in future data, it could hint for the need of an extended underlying theoretical model in the analysis.

The paper is organized as follows.  In Section~\ref{sec:ii} we introduce the dynamical DE models considered in this work and derive the time evolution of $\alpha$. Section~\ref{sec:iii} contains the description of the different probes we exploit and we highlight the main features of each observable. Section~\ref{sec:iv} details the analysis we perform and the results are presented in Sec.~\ref{sec:v}. We then discuss our results in the concluding Sec.~\ref{sec:vi}.

\section{Theoretical Models for the evolution of the fine structure constant.}
\label{sec:ii}
In this section we discuss the two broad classes of models for the evolution of the fine structure constant and present specific examples for each class, then used in the rest of the paper. In the first class, the dynamical degree of freedom providing the $\alpha$ variation is also responsible for the observationally required dark energy, while in the second class the degree of freedom is not, or only partially, responsible for the dark energy component. The observational probes are affected in different ways by these scenarios, thus leading, in principle, to constraints on DE parameters and on the coupling with electromagnetism which are specific to the particular model.

\subsection{\label{dense} Type I models: A single dynamical degree of freedom}

In this first case we assume that there is a single additional degree of freedom (typically, a scalar field)  responsible for the cosmic acceleration, and coupled to the electromagnetic sector, thus leading to the time variation of the fine structure constant $\alpha$. We consider two different models for the DE component: a phenomenological generic parametrization of the DE equation of state parameter, the Chevallier-Polarski-Linder (CPL) parametrization, and a more physically motivated Early Dark Energy (EDE) model.

\begin{itemize}
\item In the CPL model \cite{Chevallier:2000qy,Linder:2002et} the DE equation of state (EoS) is written as
\begin{equation} \label{eq:cpl}
w_{\rm CPL}(z)=w_0+w_a \frac{z}{1+z}\,,
\end{equation}
where $w_0$ is the present value of $w_{\rm CPL}$ (i.e. $w_{\rm CPL}(z=0) = w_0$) and $w_a$ is the coefficient of the time-dependent term of the EoS. 

In this model the EoS has a trend with redshift that is not intended to mimic a particular model for dark energy, but rather to allow to probe possible deviations from the $\Lambda$CDM standard paradigm without the assumption of any underlying theory. Nevertheless, we can assume that also this kind of DE is produced by a scalar field.

\item In the EDE model \cite{Doran:2006kp}, the dark energy density fraction $\Omega_{\rm EDE}(a)$ (i.e., the fraction of energy density of the DE component over the total energy density) and equation of state $w_{\rm EDE}(a)$ are parametrized in the following way
\begin{eqnarray}
\Omega_{\rm EDE}(a) &=& \frac{\od^0 - \omegae \left(1- a^{-3 w_0}\right) }{\od^0 + \Omega_{\rm m}^{0} a^{3w_0}} \nonumber \\
& &+ \omegae \left (1- a^{-3 w_0}\right) \label{eq:edeom}\\
w_{\rm EDE}(a) & =&-\frac{1}{3[1-\Omega_{\rm EDE}]} \frac{d\ln\Omega_{\rm EDE}}{d\ln a} \nonumber \\
&& + \frac{a_{eq}}{3(a + a_{eq})} 
\label{eq:edew} 
\end{eqnarray}
where $\aeq$ is the scale factor at matter-radiation equality and $\od^0$ and $\Omega_{\rm m}^{0}$ are the current dark energy and matter density, respectively. A flat universe is assumed and the present value for the equation of state is obtained demanding $w(a=1)=w_0$. The energy density $\od(a)$ has a scaling behaviour evolving with time and going to a finite constant $\ome$ in the past.

In this case the EoS follows the behaviour of the dominant component at each cosmic time; $w_{\rm EDE}\approx1/3$ during radiation domination, $w_{\rm EDE}\approx0$ during matter domination, and $w_{\rm EDE}\approx-1$ in recent times, as in a cosmological constant era.  We add dark energy perturbations as in \cite{Calabrese:2010uf} but we fix the clustering parameters to the values expected in the case of a scalar field.
\end{itemize}

In these models the dynamical scalar fields are expected to be naturally coupled to the rest of the theory, unless a (still unknown) symmetry suppresses this coupling \cite{Carroll}.  We assume that this is the case for our toy models too, and, following the line of \cite{Calabrese:2011nf}, we want to study the coupling of the dark energy degree of freedom with the electromagnetic field.

The coupling between the scalar field, $\phi$, and electromagnetism stems from a gauge kinetic function $B_F(\phi)$
\begin{equation}
{\cal L}_{\phi F} = - \frac{1}{4} B_F(\phi) F_{\mu\nu}F^{\mu\nu}
\end{equation}
which, to a good approximation, can be assumed linear \cite{Nunes,Couplings1}, 
\begin{equation}
B_F(\phi) = 1- \zeta \sqrt{8\pi G} (\phi-\phi_0)\ .
\end{equation}

This form of the gauge kinetic function can be seen as the first term of a Taylor expansion, which is indeed a good approximation for a slowly varying field at low redshifts, as the low-redshift constraints on couplings, obtained both directly from astrophysical measurements and through local tests of equivalence principle violations, are quite tight. For the latter category we can refer to the conservative constraint \cite{Pospelov,Dvali}
\begin{equation}
|\zeta_{local}|<10^{-3}\,.\label{localzeta}
\end{equation}
In \cite{Calabrese:2011nf}, the authors obtained an independent few-percent constraint on this coupling using CMB and large-scale structure data in combination with direct measurements of the expansion rate of the universe. 

With these assumptions, the evolution of $\alpha$ is given by 
\begin{equation}
\frac{\Delta \alpha}{\alpha} \equiv \frac{\alpha-\alpha_0}{\alpha_0} =
\zeta \sqrt{8\pi G} (\phi-\phi_0) \,,
\end{equation}
and, since the evolution of the putative scalar field can be expressed in terms of the dark energy
properties $\Omega_\phi$ and $w$ as \cite{Couplings1,Couplings2}
\begin{equation}
w = -1 + \frac{(\sqrt{8\pi G}\phi')^2}{3 \Omega_\phi} \,,
\end{equation}
where the prime denotes the derivative with respect to the logarithm of the scale factor,
we finally obtain the following explicit relation for the evolution of the fine
structure constant in this dynamical dark energy class of models
\begin{equation} \label{eq:dalfa}
\frac{\Delta\alpha}{\alpha}(z) =\zeta \int_0^{z}\sqrt{3\Omega_\phi(z)\left[1+w(z)\right]}\frac{dz'}{1+z'}\,.
\end{equation}

As expected, in this class of models the magnitude of the $\alpha$ variation is controlled by the strength of the coupling $\zeta$. We also note that these two equations can be phenomenologically generalized to the case of phantom equations of state, by simply switching the sign of the $(1+w)$ term \cite{Vielzeuf:2013aja}.

Here $\Omega_\phi(z)$ is the fraction of energy density provided by the scalar field, thus it corresponds to Eq.(\ref{eq:edeom}) in the EDE case, while for the CPL parametrization it's easily found to be
\begin{equation}
\Omega_{\rm CPL}(z)=\frac{\Omega_{CPL}^0}{\Omega_{\rm CPL}^0+\Omega_{\rm m}^0(1+z)^{-3(w_0+w_a)}e^{(3w_az/1+z)}}\,.
\end{equation}
where $\Omega_{\rm m}^0$ and $\Omega_{\rm CPL}^0$ are, respectively, the present time energy densities of matter and DE.

\subsection{\label{light} Type II models: Independent degrees of freedom}

In this scenario the degree of freedom responsible for the $\alpha$ variation does not provide the dark energy, or at least is constrained to provide only a fraction of it by current observations. One effectively has a $\Lambda$CDM model with an additional (often phenomenological) degree of freedom accounting for the $\alpha$ variation.

In this case the direct link between varying couplings and dark energy discussed above is also lost. Nevertheless, it is possible to observationally infer that a given $\alpha$ variation is \emph{not} due to a Type I model, as such an assumption could lead to consequences that can be observationally ruled out. This possibility has already been discussed in \cite{Vielzeuf:2012zd}. Here we will discuss this class in a slightly different context.

The simplest toy model of this kind is the Bekenstein-Sandvik-Barrow-Magueijo (BSBM) model \cite{Sandvik:2001rv}. These theories require some fine-tuning, even to fit purely temporal $\alpha$ variations as that of \cite{Murphy}, but for our purposes they are useful for parametrizing the biases introduced in cosmological parameter estimations if there is an $\alpha$ variation which is neglected in the analysis. For the $\alpha$ variation itself we can, to a good approximation, assume a simple one-parameter ($\xi$) evolution, like
\begin{equation}\label{eq:bsbm}
\frac{\Delta\alpha}{\alpha}=-4\xi\ln{(1+z)} \,.
\end{equation}

An alternative example of this class is provided by the string-theory inspired runaway dilaton scenario \cite{Damour:2002mi}, where the $\alpha$ evolution is also relatively simple.

\section{Observational Probes}
\label{sec:iii}
In this section we characterize the different observables we will use in our analysis.

\subsection{Supernovae Type Ia data}

Type Ia Supernovae are a particular class of Supernovae, providing bright, standardizable candles, and constraining cosmic acceleration through  the Hubble diagram. At present, they are the most effective and mature probe of dark energy.

Moreover, as the SN peak luminosity ($L_{peak}$) depends on photon diffusion time, which in turn depends on $\alpha$ through the opacity, the $\alpha$ variation could affect $L_{peak}$ \cite{Chiba:2003hz}. The key mechanism is the energy deposition rate in the decay chain ${}^{56}Ni\rightarrow{}^{56}Co\rightarrow{}^{56}Fe$. This leads to
\begin{equation}
\frac{\Delta L_{peak}}{L_{peak}}\sim\,- 0.94\, \frac{\Delta \alpha}{\alpha}
\end{equation}
which corresponds to
\begin{equation}
\frac{\Delta \alpha}{\alpha}\sim0.98\, \Delta M
\end{equation}
where $\Delta M=M-M_0$ with $M$ the absolute magnitude at peak, and the subscript ``$0$" indicates we are not accounting for the $\alpha$ variation. 

Decreasing alpha decreases the opacity, allowing photons to escape faster, thus increasing $L_{peak}$. This can be trivially translated to a change in the distance modulus $\mu=m-M$, with $m$ the apparent magnitude as
\begin{equation}\label{eq:snalfa}
 \mu(z)=m-M=m-(M_0+\Delta M)=\mu_0(z)-\frac{1}{0.98}\frac{\Delta\alpha}{\alpha}(z)
\end{equation}
where $\mu_0(z)=5 \log_{10}(d_L(z))+25$ is function of the luminosity distance
\begin{equation}
 d_L(z)=\frac{1+z}{H_0}\int_0^z{\frac{dz}{E(z)}} \ .
\end{equation}
The $E(z)=H(z)/H_0$ expression encodes the chosen dark energy model. 

We build the SN datasets following the procedure presented in \cite{Cardone:2012vd}.
We use Euclid specifications \cite{Laureijs:2011mu,Hook:2012xk} to forecast a SN survey at low-intermediate $z$, containing $1700$ supernovae uniformly distributed in the redshift range $0.75<z<1.5$.

\subsection{Quasar absorption systems data}

The frequencies of narrow metal absorption lines in quasar absorption systems are sensitive to $\alpha$ \cite{Bahcall}, and the different transitions have different sensitivities. Observationally, one expects relative velocity shifts between transitions in a given absorber, in a single spectrum, if $\alpha$ does vary; this comparison can therefore be used to obtain measurements of $\alpha$ in these absorption systems. Indeed a survey able to observe quasar absorption lines at different redshifts is able to reconstruct the variation of $\alpha$ with respect to the present value and to provide a dataset corresponding to the left side of Eq.~(\ref{eq:dalfa}).

Currently, there is controversial evidence \cite{Webb} for a space-time variation of $\alpha$ at the level of a few parts per million, roughly in the redhsift range $1<z<4$. Part of the uncertainty in these results stems from the fact that the large samples of spectra being used have been gathered for other purposes and are therefore inhomogeneous, and may be vulnerable to systematic errors which are difficult to quantify. An ongoing dedicated VLT-UVES Large Program is trying to clarify this issue \cite{LP1,LP2}, but the ultimate solution is to use high-resolution ultra-stable spectrographs, for which these measurements are a key science driver.

For representative future datasets we use the baseline (conservative) case discussed in \cite{Amendola:2011qp}. We consider the European Extremely Large Telescope (E-ELT) equipped with a high-resolution, ultra-stable spectrograph (ELT-HIRES), for which the COsmic Dynamics Experiment (CODEX) Phase A study \cite{codex} provides a baseline reference. We assume uniformly distributed measurements in the redshift range $0.5<z<4.0$, with an error $\sigma_\alpha=10^{-7}$.

\subsection{Redshift-drift data}

QSO observations can be also used to constrain DE models through the so called redshift-drift of these sources \cite{Sandage,Loeb:1998bu}. The redshift-drift is the change of the redshift due to the expansion of the universe between two observations of the same distant source spectrum, repeated after a given amount of (terrestrial) years. The required time lapse depends on the instrument used (and specifically on its calibration stability) but is typically of the order of a decade with next-generation facilities.

With this kind of observations one can exploit distant astrophysical sources as a probe of the expansion of the universe in a model independent way \cite{Pasquini,Corasaniti:2007bg,Quercellini:2010zr}. As pointed out in \cite{Vielzeuf:2012zd,Martinelli:2012vq} QSO are the ideal astrophysical objects to observe the redshift variation $\Delta z$ between two observations. This $\Delta z$ can be translated to a spectroscopic velocity $\Delta v=c\Delta z/(1+z)$ and connected to cosmological quantities through the relation
\begin{equation}
 \frac{\Delta v}{c}=H_0\Delta t\left[1-\frac{E(z)}{1+z}\right],
\end{equation}
where $c$ is the speed of light and $\Delta t$ is the time interval between two observations of the same astrophysical source.

A CODEX-like spectrograph will have the ability to detect the cosmological redshift-drift in the Lyman $\alpha$ absorption lines of distant (2 $<z<$ 5) QSOs, even though this is a very small signal. The E-ELT can decisively detect the redshift variation with a 4000 hours of integration in a period of $\Delta t=20$ years \cite{Liske:2008ph}. These may be complemented by measurements at other redshifts using SKA \cite{SKA1,SKA2}.

According to Monte Carlo simulations of the CODEX Phase A study \cite{codex}, the error on the measured spectroscopic velocity shift $\Delta v$ that can be expressed as:
\begin{equation}\label{eq:error}
 \sigma_{\Delta v}=1.35\ \frac{2370}{S/N}\ \sqrt{\frac{30}{N_{\rm QSO}}}\ \left(\frac{5}{1+z_{\rm QSO}}\right)^x\ cm\ s^{-1},
\end{equation}
where $S/N$ is the signal to noise ratio, $N_{\rm QSO}$ the number of observed quasars, $z_{\rm QSO}$ their redshift and the exponent $x$ is equal to $1.7$ when $z\leq4$, while it becomes $0.9$ beyond that redshift.

Therefore, we can forecast a redshift-drift dataset where the error bars are computed using Eq.(\ref{eq:error}), with $S/N=3000$ and a number of QSO $N_{\rm QSO}=30$ is assumed to be uniformly distributed among the following redshift bins $z_{\rm QSO}=[2.0,2.8,3.5,4.2,5.0]$.

\subsection{Weak lensing data}

Weak gravitational lensing of distant galaxies is a powerful observable to probe the geometry of the universe and to map the dark matter distribution.
We describe the distortion of the images of distant galaxies through the tensor~\cite{Bartelmann:1999yn}\\
\begin{center}
\(\psi_{ij}=\left(
\begin{array}{cc}
-\kappa-\gamma_1 & -\gamma_2 \\
-\gamma_2 & -\kappa+\gamma_2 \\
\end{array}\right)\)
\end{center}
where $\kappa$ is the convergence field and $\gamma=\gamma_1+i\gamma_2$ is the complex shear field. We can rewrite these quantities as a function of the projected Newtonian potentials $\psi_{,ij}$ 
$$\gamma=\frac{1}{2}(\psi_{,11}-\psi_{,22})+i\psi_{,12} \ ,$$
$$\kappa=\frac{1}{2}(\psi_{,11}-\psi_{,22}) \ $$ 
where the commas indicate the derivatives with respect to the directions transverse to the line of sight and the projected potentials are given by
$\psi_{,ij}=-(1/2)\int g(z)(\Psi_{,ij}+\Phi_{,ij})dz$, i.e. integrating the gravitational potentials with the lensing kernel
$$g(z)=\int dz^\prime\frac{n(z^\prime)r(z,z^\prime)}{r(0,z^\prime)} \ $$
with $n(z)$ the galaxy redshift distribution and $r$ the comoving distance
\begin{equation}
 r(z,z')=\int_z^{z'}{\frac{dz''}{E(z'')}}.
\end{equation}

We can define the convergence power spectra in a given redshift bin in the following way
\begin{equation}\label{pidielle}
 P_{ij}(\ell)=H_0^3\int_0^\infty\frac{dz}{E(z)}W_i(z)W_j(z)P_{\rm NL}[P_{\rm L}\left(\frac{H_0\ell}{r(z)},z\right)]
\end{equation}
where $P_{\rm NL}$ is the non-linear matter power spectrum at redshift
$z$, obtained correcting the linear one $P_{\rm L}$. $W(z)$ is a weighting function
\begin{equation}\label{weight}
    W_i(z)=\frac{3}{2}\Omega_m(1+z)\int_{z_i}^{z_{i+1}}dz^\prime\frac{n_i(z^\prime)r(z,z^\prime)}{r(0,z^\prime)}
\end{equation}
with subscripts $i$ and $j$ indicating the redshift bin.

The observed power spectra are affected mainly by systematic uncertainties arising from the intrinsic ellipticity of galaxies $\gamma^2_{\rm rms}$. These uncertainties can be reduced averaging over a large number of sources. The observed convergence power spectra will be hence
\begin{equation}\label{obsconv}
    C_{ij}=P_{ij}+\delta_{ij}\gamma^2_{\rm rms}\tilde{n}_j^{-1}
\end{equation}
where $\tilde{n}_j$ is the number of sources per steradian in the $j-th$ bin.

In this paper we simulate a weak lensing dataset according to the specifications expected for the Euclid survey \cite{Laureijs:2011mu}: the mission will observe $n_g \simeq 30 \ {\rm gal/arcmin^2}$ over an area $\Omega = 15000 \ {\rm deg^2}$, corresponding to a sky fraction $f_{sky} \sim 33\%$. The large galaxy number density and the wide area observed will allow Euclid to provide us with a tomographic reconstruction of the weak lensing signal. We therefore divide the redshift space in 10 bins, chosen in such a way to have the same fraction of the total observed galaxies in each one (see Table \ref{tab:bins}). Using these specifications we build the $\ell$-by-$\ell$ convergence power spectrum and the $1 \sigma$ uncertainties, computed as \cite{fdebe011,Cooray:1999rv}\,
\begin{equation}
\label{sigmaconv}
\sigma_{\ell}=\sqrt{\frac{2}{(2\ell+1)f_{sky}}}\left(P(\ell)+\frac{\gamma_{rms}^2}{n_{gal}}\right)~.
\end{equation}

\begin{table}
\begin{center}
\begin{tabular}{c|c|c|c}
\hline
bin & $z$           & bin  & $z$ \\
\hline
$1$ & $0-0.496$     & $6$  & $1.031-1.163$ \\
$2$ & $0.496-0.654$ & $7$  & $1.163-1.311$ \\
$3$ & $0.654-0.784$ & $8$  & $1.311-1.502$ \\
$4$ & $0.784-0.907$ & $9$  & $1.502-1.782$ \\
$5$ & $0.907-1.031$ & $10$ & $1.782-5.000$ \\
\hline
\end{tabular}
\caption{Euclid redshift bins considered in this analysis. The redshift range of every bin is chosen in such a way that each bin contains 10\% of the galaxies observed by the survey.}
\label{tab:bins}
\end{center}
\end{table}

\subsection{Atomic clocks bounds}

In models where the same dynamical degree of freedom is responsible for both the dark energy and the variation of $\alpha$, at redshift $z=0$ the atomic clock bounds \cite{Rosenband} will always give a constraint on the combination of a fundamental physics parameter (e.g. the coupling of the field, which is obtained by the Equivalence Principle violation) and a cosmological parameter (usually the dark energy equation of state $w_0$, although depending on the model other parameters may be involved too). For the models in subsection \ref{dense}, we have
\begin{equation}
\sqrt{3\Omega_{\phi 0}(1+w_0)} H_0\zeta = (-1.6\pm2.3)\times10^{-17} yr^{-1}\,,
\label{eq:clocks}
\end{equation}
and there will be analogous relations for the other models. In some cases it may be possible to set such a bound at non-zero redshifts too.

For \ref{light}-like models Eq.~(\ref{eq:clocks}) simplifies to 
\begin{equation}
4 H_0\xi = (-1.6\pm2.3)\times10^{-17} yr^{-1}\,.
\label{eq:clocks2}
\end{equation}

\section{Analysis}
\label{sec:iv}

The cosmological parameters that we sample can be divided in ``standard parameters'' quantifying the content of the universe and the power spectrum of primordial scalar perturbations, \{$\Omega_{b}h^2$,$\Omega_{c}h^2$,$\Omega_\Lambda$,$n_s$,$A_s$\}, peculiar DE parameters characterizing different parametrizations, \{$w_0$, $w_a$\} for the CPL case and \{$w_0$, $\Omega_e$\} for EDE, and the coupling $\zeta$ ($\xi$ for the BSBM model). 

We build simulated datasets assuming a fiducial cosmology given by the observations of the WMAP satellite after 9 years of data \cite{Hinshaw:2012aka} for the standard parameters: the baryon and cold dark matter densities, $\Omega_{b}h^2$ and $\Omega_{c}h^2$, the amount of energy density given by dark energy at the present time $\Omega_\Lambda$, the optical depth to reionization $\tau$, the scalar spectral index $n_s$ and the overall normalization of the spectrum $A_s$ (see Table \ref{tab:fiducial}). We fix the DE parameters in such a way to mimic the $\Lambda$CDM expansion (i.e. $w_0=-1,\ w_a=0$ in the CPL case and $w_0=-1,\ \Omega_e=0$ for EDE) and a vanishing coupling $\zeta=0$. In all the models and analysis we require spatial flatness of the universe. Basically, this fiducial set of parameters (Set1 in Table~\ref{tab:fid_nost}) represents the standard $\Lambda$CDM cosmology as measured by WMAP-9.
\begin{table}[h]
\begin{center}
\begin{tabular}{c|c|c|c|c|c}
\hline
$\Omega_bh^2 $ & $\Omega_ch^2 $ & $\Omega_\Lambda$ & $\tau$ & $n_s $ & $A_s$\\
\hline
$\ 0.02264\ $ & $\ 0.1138\ $ & $\ 0.722\ $ & $\ 0.089\ $ & $\ 0.972\ $ & $\ 2.40\cdot 10^{-9}\ $\\
\hline
\end{tabular}
\caption{Fiducial values for the six standard $\Lambda$CDM cosmological parameters, corresponding to the marginalized best fit values of the WMAP-9 years analysis.}
\label{tab:fiducial}
\end{center}
\end{table}
\begin{table}[h]
\begin{center}
\begin{tabular}{c|c|c|c|c|c}
\hline
Fiducial & $w_0$ & $w_a$ & $\Omega_e$ & $\zeta$ & $\xi$ \\
\hline
Set1     & $-1$    & $0$   & $0$        & $0$                  & $-$  \\
Set2     & $-0.95$ & $0$   & $-$        & $-3\times10^{-5}$    & $-$  \\
Set3     & $-0.95$ & $-$   & $0.02$     & $-2\times10^{-5}$    & $-$  \\
Set4     & $-$     & $-$   & $-$        & $-$                  & $5\times10^{-8}$  \\
\hline
\end{tabular}
\caption{Fiducial values for the DE parameters and couplings used in the different analyses.}
\label{tab:fid_nost}
\end{center}
\end{table}

We also build simulated datasets with a non vanishing variation of $\alpha$ assuming the same value of Table \ref{tab:fiducial} for the standard parameters, but different values for the ones involved in the $\alpha$ variation, listed in Table~\ref{tab:fid_nost}. In order to produce an evolving $\alpha$, DE parameters must depart from the standard $\Lambda$CDM scenario, nevertheless we assume fiducial model values compatible with presently available constraints \cite{Ade:2013lta,Pospelov,Dvali,pettorino,act} . In particular, for the CPL case we assume $w_0=-0.95,\ w_a=0$ and a coupling $\zeta=-3\times10^{-5}$ (Set2).  For the EDE case we choose a dark energy described by $w_0=-0.95,\ \Omega_e=0.02$ and a coupling $\zeta=-2\times10^{-5}$ (Set3). We exploit these last two datasets to constrain the DE parameters beyond the standard $\Lambda$CDM model and in order to investigate the possible bias on cosmological parameters introduced if we neglect the variation of $\alpha$ in the analysis.

In the BSBM framework instead we only use one fiducial model (Set4) generating a non vanishing $\Delta\alpha/\alpha$ with a coupling $\xi=5\times10^{-8}$, in order to inquire how the possible presence of a scalar field not driving the accelerated expansion, but coupled with $\alpha$, can bias the recovered cosmological parameters. In this case the DE parameters are fixed to the $\Lambda$CDM values as we assume that the background expansion is not affected by this scalar field.

\begin{figure}[htb!]
\begin{center}
\includegraphics[width=\columnwidth]{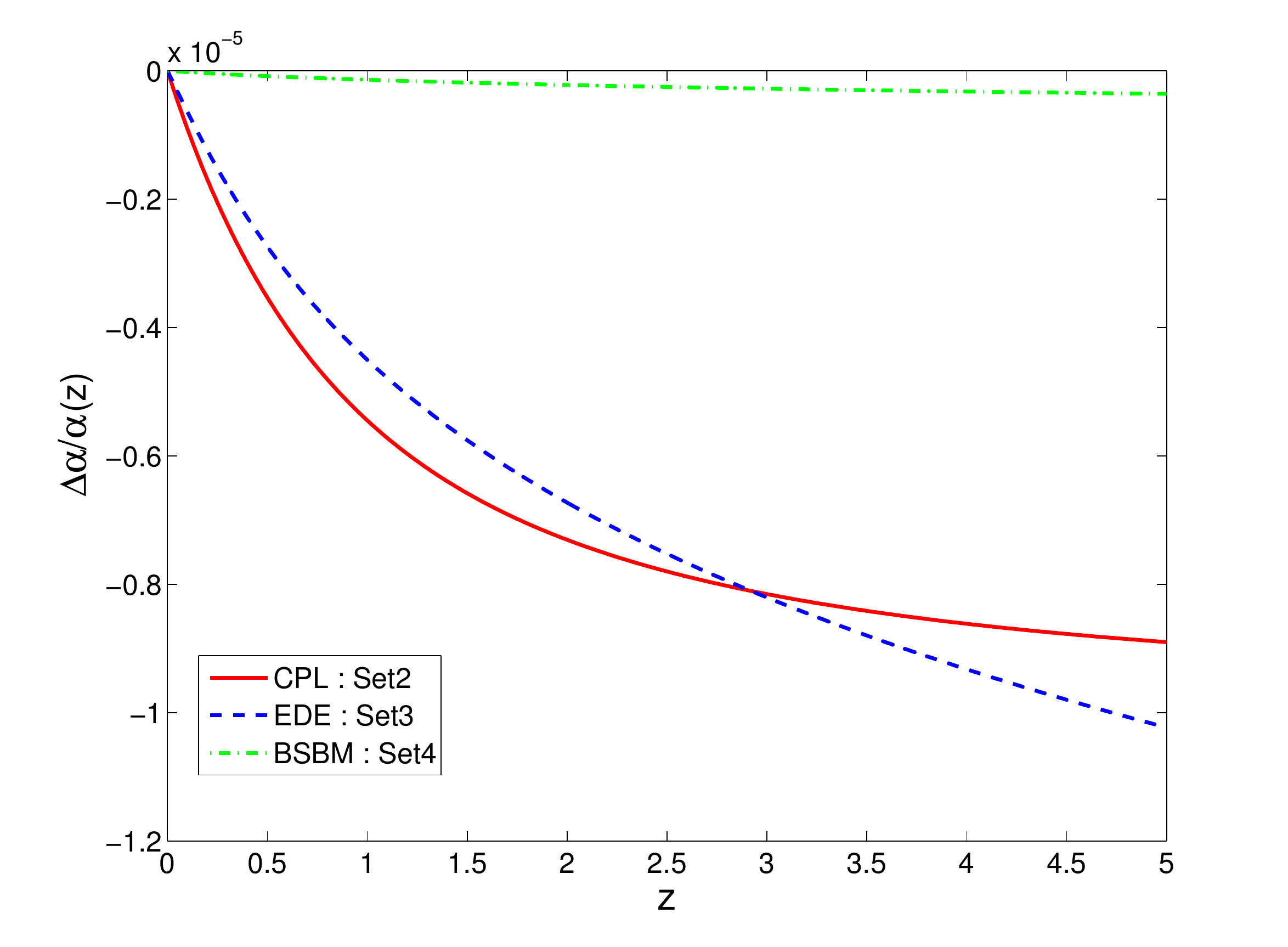}\\
\includegraphics[width=\columnwidth]{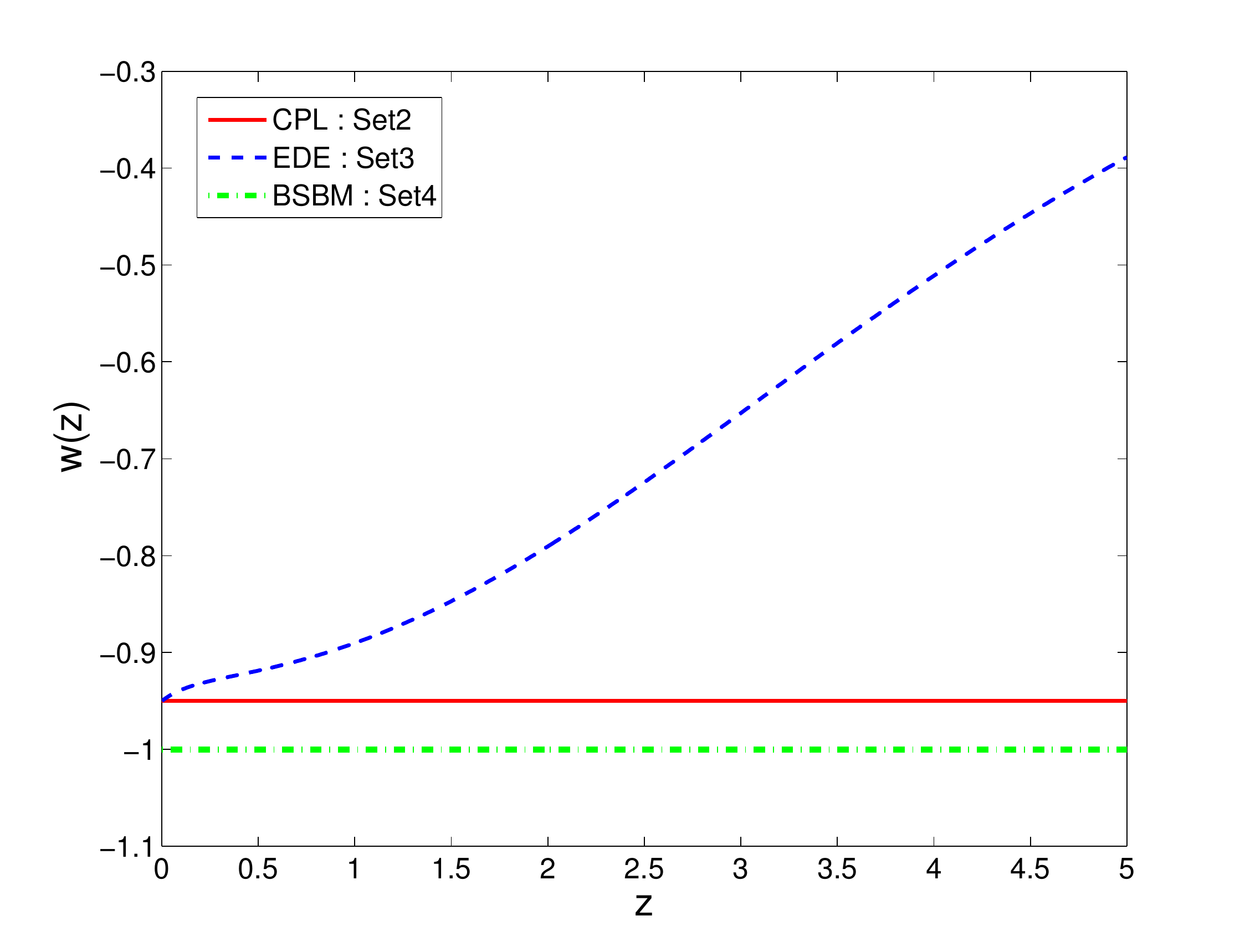}\\
\caption{Top panel: Evolution with redshift of \dalfa in the CPL (red solid line), EDE (blue dashed line) and BSBM (green dash-dotted line) parametrizations using the fiducial cosmology in Table~\ref{tab:fid_nost}. Bottom panel: corresponding variation in the DE equation of state. }
\label{fig:dalfa}
\end{center}
\end{figure}

We show in Fig.~\ref{fig:dalfa} the resulting time variation of $\alpha$ (top panel) and the corresponding EoS (bottom panel) for the non standard scenarios defined by Table~\ref{tab:fid_nost}. 

 In this work we rely on a MCMC technique to sample the parameter space and we use a modified version of the publicly available package \texttt{cosmomc} \cite{Lewis:2002ah} with a convergence diagnostic using the Gelman and Rubin statistics. We assume flat priors on the sampled parameters.

\section{Results}
\label{sec:v}
In this section we present the most interesting results we obtained, discussing the impact of different observables on the constraints. The complete set of constraints, resulting from using different combinations of probes, is reported in the Appendix \ref{sec:app}.

\subsection{Vanishing $\Delta\alpha/\alpha$}

As stated in the previous section, the first investigation we carry out deals with vanishing \dalfa mock datasets. We consider different combinations of the probes introduced in Section~\ref{sec:iii} and discuss the main features obtained by this analysis, exploring how the main geometrical probes (WL and SN) affect constraints on DE parameters and on the coupling $\zeta$.

\begin{figure}[t!]
\begin{center}
\includegraphics[width=\columnwidth]{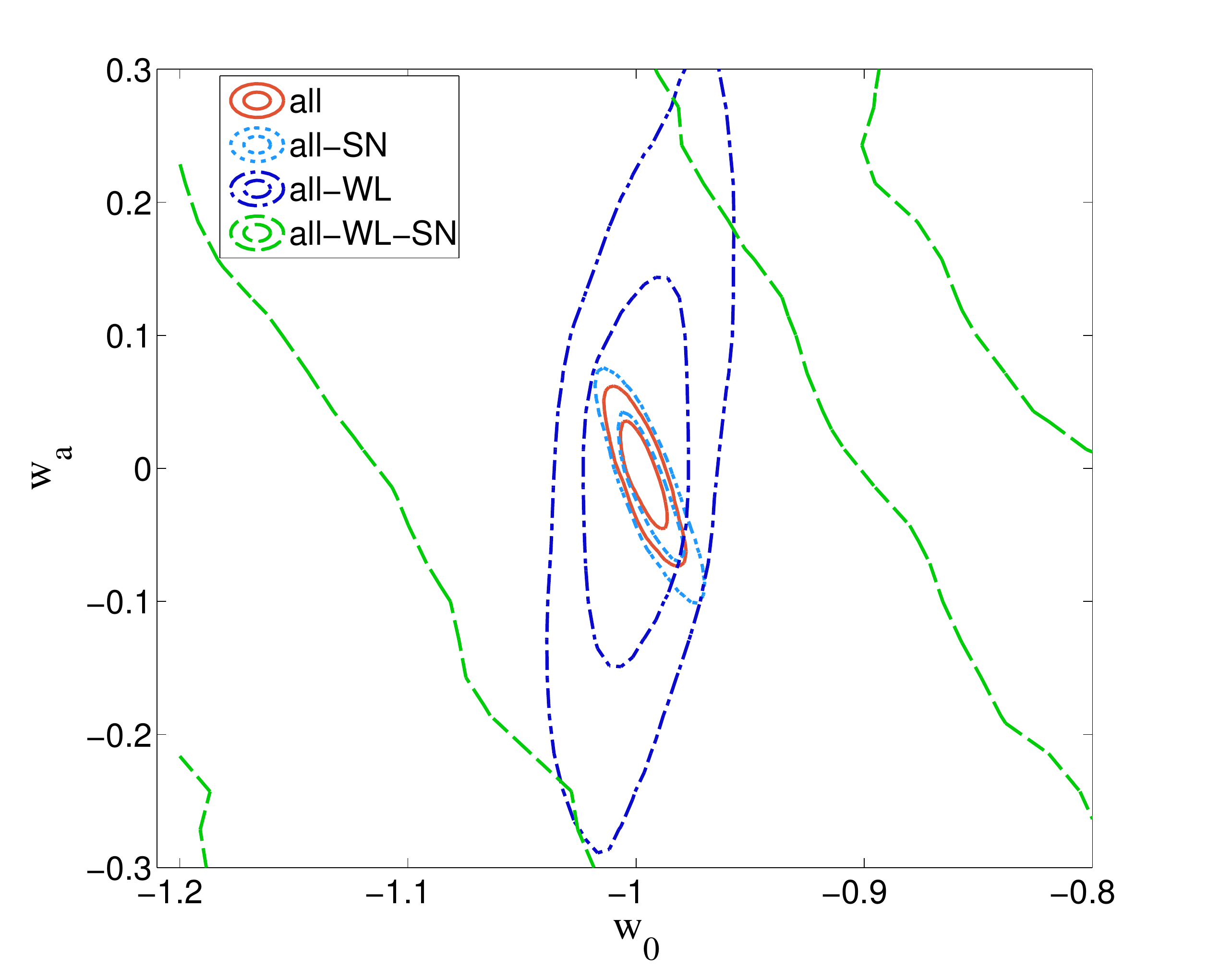}
\caption{2-dimensional contours at 68\% and 95\% confidence levels for the $w_0$-$w_a$ parameters. The solid red contours show the combination of all observables; dotted cyan curves describe the degradation of the constraints when removing SN; blue dot-dashed contours broaden because of the exclusion of WL; the green dashed regions are obtained removing both WL and SN measurements. }
\label{fig:w0wa}
\end{center}
\end{figure}

We first report the results for the CPL model. In Fig.~\ref{fig:w0wa} we can notice how the Euclid survey will greatly narrow the allowed parameter space for the EoS parameters $w_0$ and $w_a$, mainly thanks to the combination of the SN and WL measurements. When we consider all datasets we get $\sigma(w_0) = 0.007$ and $\sigma(w_a)=0.03$. 
\begin{figure}[t!]
\begin{center}
\includegraphics[width=\columnwidth]{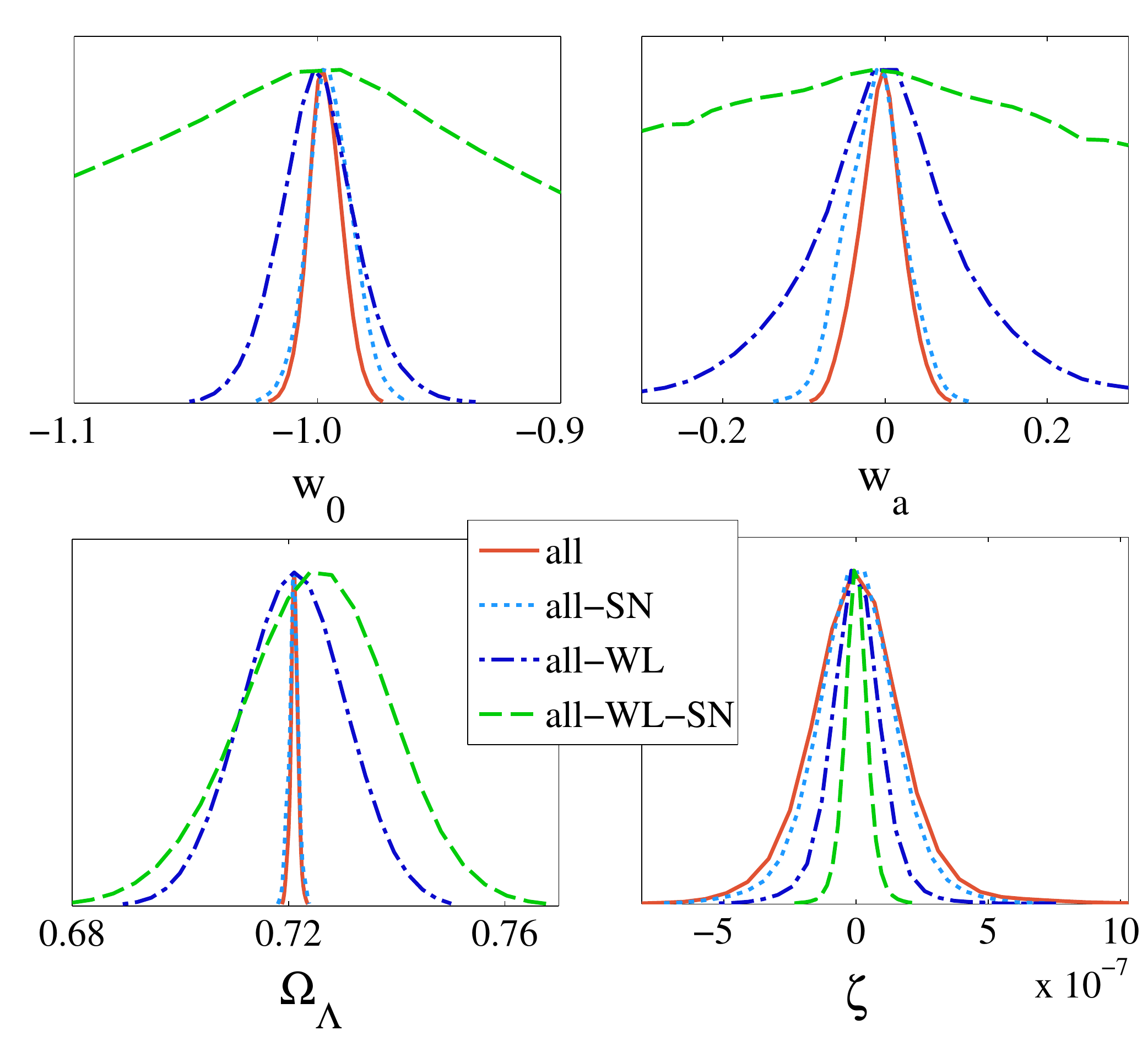}
\caption{Marginalized 1-dimensional posterior distributions for the DE parameters $w_0$, $w_a$, $\Omega_\Lambda$ and the coupling $\zeta$, for different combinations of probes.}
\label{fig:zetaCPL}
\end{center}
\end{figure}

\begin{figure}[b!]
\begin{center}
\includegraphics[width=\columnwidth]{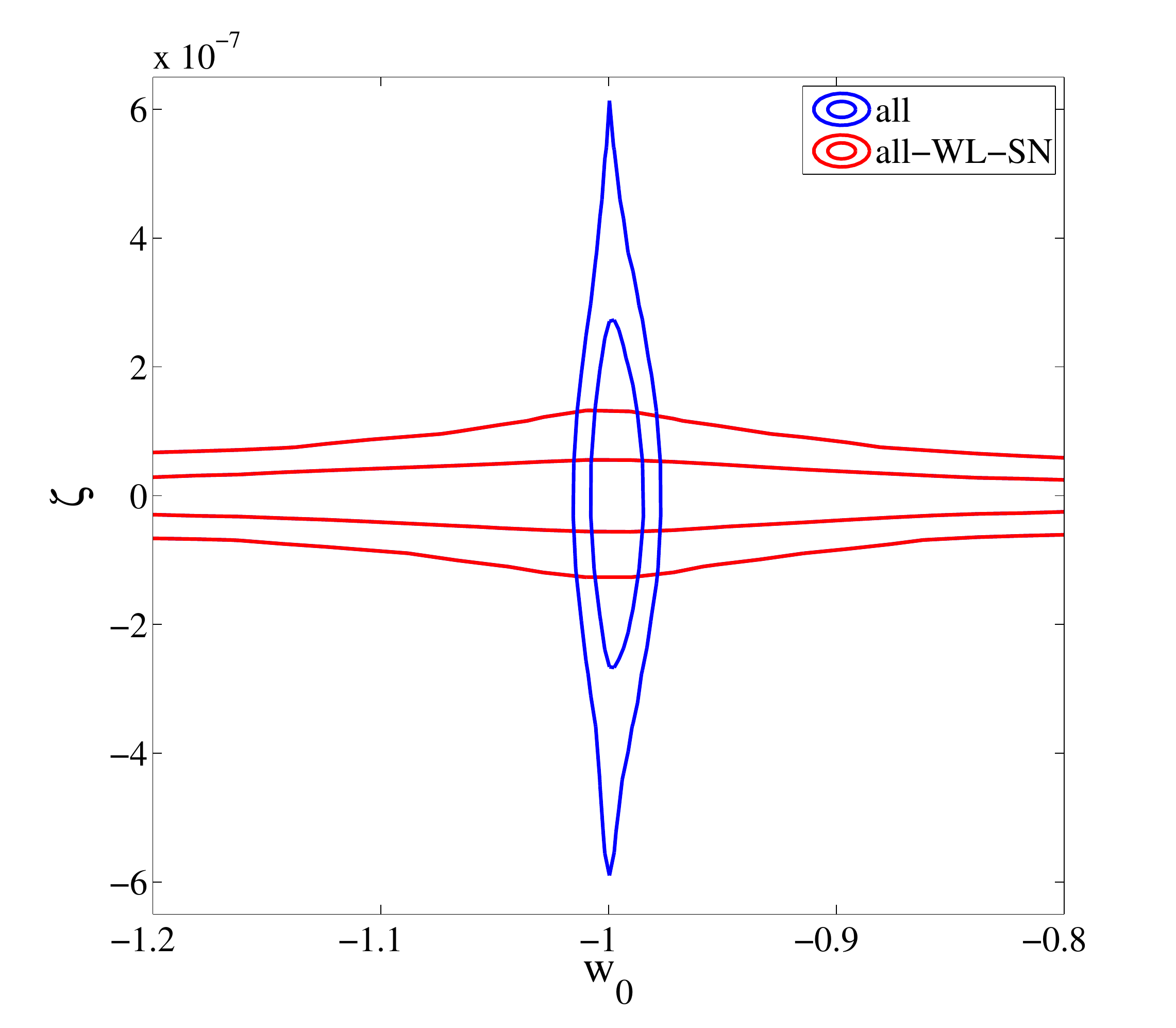}\\
\includegraphics[width=\columnwidth]{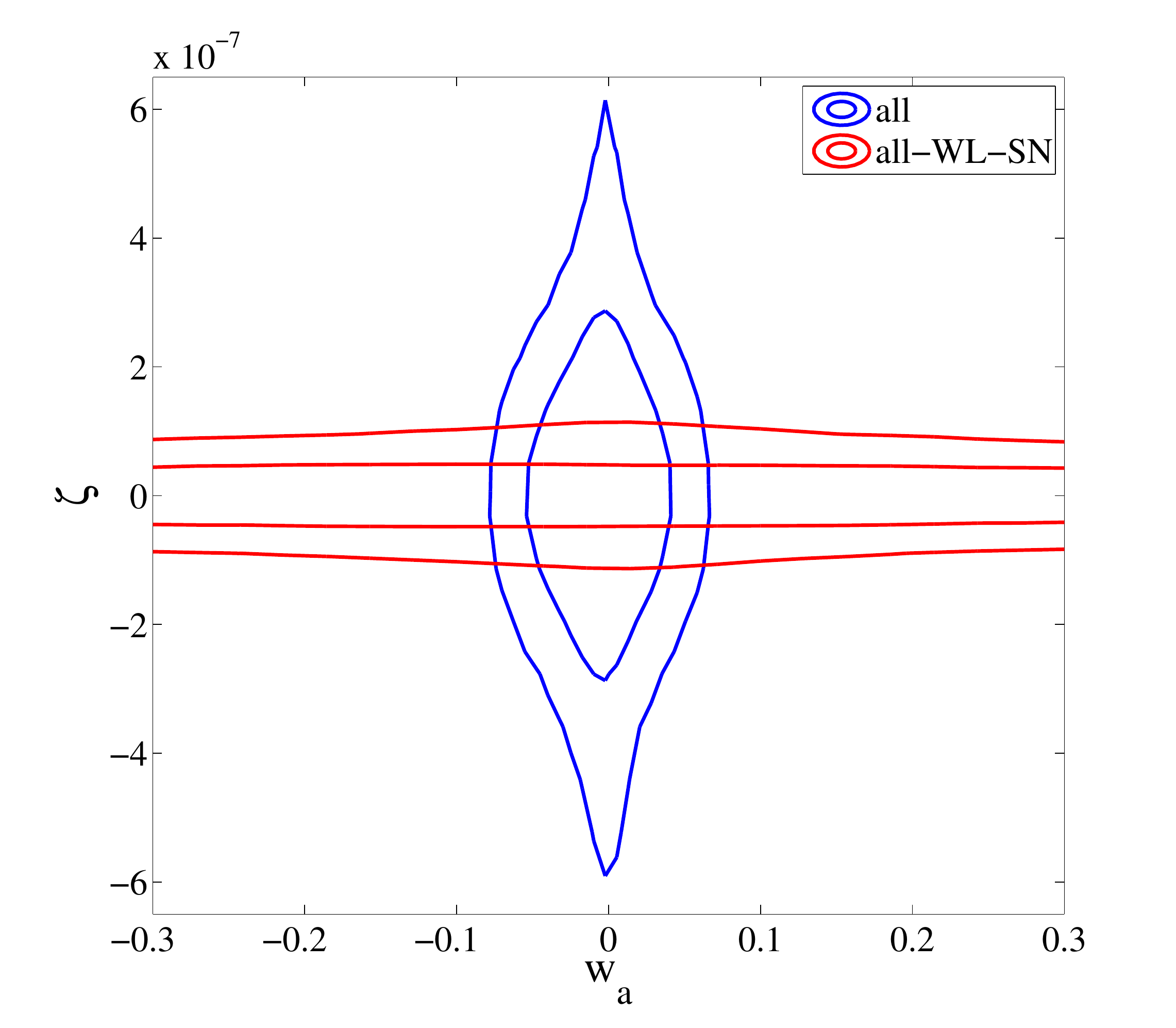}
\caption{2-dimensional contours at 68\% and 95\% confidence levels showing $\zeta$ versus $w_0$/$w_a$ with (closed blue contours) and without (open red contours) the inclusion of WL and SN observations.}
\label{fig:trendCPL}
\end{center}
\end{figure}

The constraints on the coupling parameter are instead puzzling at a first look (see panel 4 in Fig.~\ref{fig:zetaCPL}), as the use of the Euclid observations loosens the bounds on $\zeta$. 
This result is however easily explained considering the chosen fiducial cosmological model. Eq.(\ref{eq:dalfa}) in fact implies that a vanishing \dalfa can be obtained in two ways: either $\zeta=0$ and/or $w(z)=-1$. This leads to the fact that when $w_0$ and $w_a$ are poorly constrained (i.e. when WL and SN are removed from the analysis) the QSO forecasted measurements require a coupling $\zeta$ close to zero. On the contrary when WL and SN impose tight independent constraints on DE parameters and the recovered $w(z)$ is close to $-1$, a larger range of $\zeta$ values is in agreement with the QSO measurements.
We can interpret this result considering that, as our chosen fiducial cosmology is the standard $\Lambda$CDM universe, our probes tightly constrain the Dark Energy to be close to a cosmological constant, thus a non dynamical field (or one rolling down the potential extremely slowly), and therefore a vanishing \dalfa is reproduced for every choice of the coupling.
This effect is displayed in Fig.~\ref{fig:zetaCPL} where we report the recovered 1-dimensional posterior distributions for the coupling and the DE parameters. The solid red curves show the combination of all observables with very tight constraints on DE parameters and the larger distribution for $\zeta$; the dotted cyan curves are obtained removing SN, the constraints on $w_0$-$w_a$ are slightly broader and the coupling is slightly better constrained; the blue dot-dashed lines exclude WL: DE parameters are still measured by SN but the constraints are largely broadened allowing for a tighter measurement of $\zeta$; the green dashed lines show the constraints on parameters when removing both WL and SN: in this case we get the most stringent constraint on the coupling because of the unmeasured $w_0$-$w_a$ parameters. 
In Fig.~\ref{fig:trendCPL} we show the 2-dimensional contours at 68\% and 95\% confidence levels in the $\zeta$-$w_0$ and $\zeta$-$w_a$ planes only for the two extreme cases: the combination of all probes and the analysis excluding WL and SN. Again we can see that when DE parameters are constrained thanks to WL and SN, the coupling can lie in a larger region, while it is tightly constrained when loose bounds on $w_0$-$w_a$ are obtained. 

\begin{figure}[htb!]
\begin{center}
\includegraphics[width=\columnwidth]{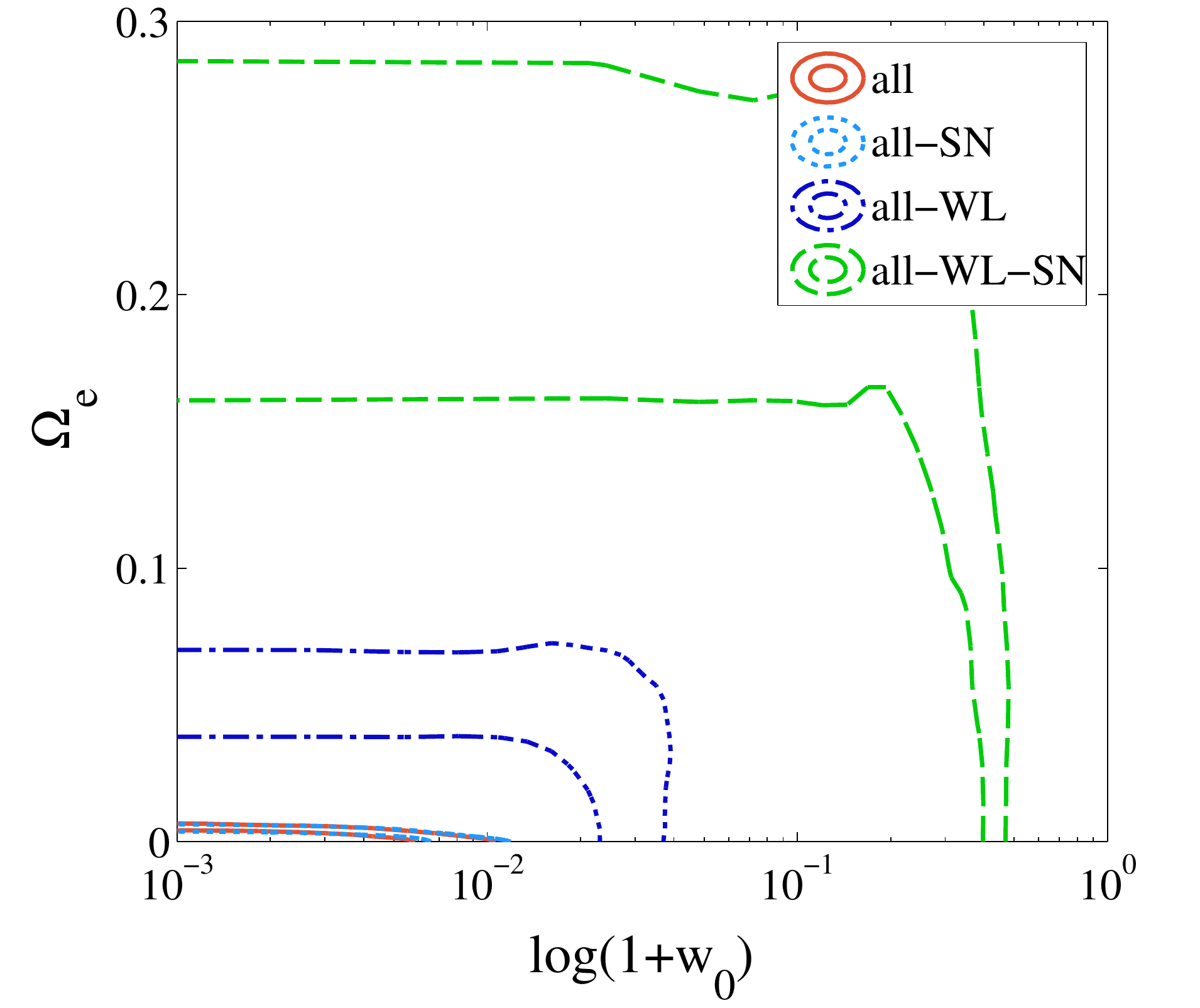}
\caption{Same as Fig.~\ref{fig:w0wa} for the EDE parameters $w_0$-$\Omega_{\rm e}$. Here we plot $log(1+w_0)$ to better show the $w_0\sim-1$ region.}
\label{fig:w0oe}
\end{center}
\end{figure}

\begin{figure}[htb!]
\begin{flushright}
\includegraphics[width=\columnwidth]{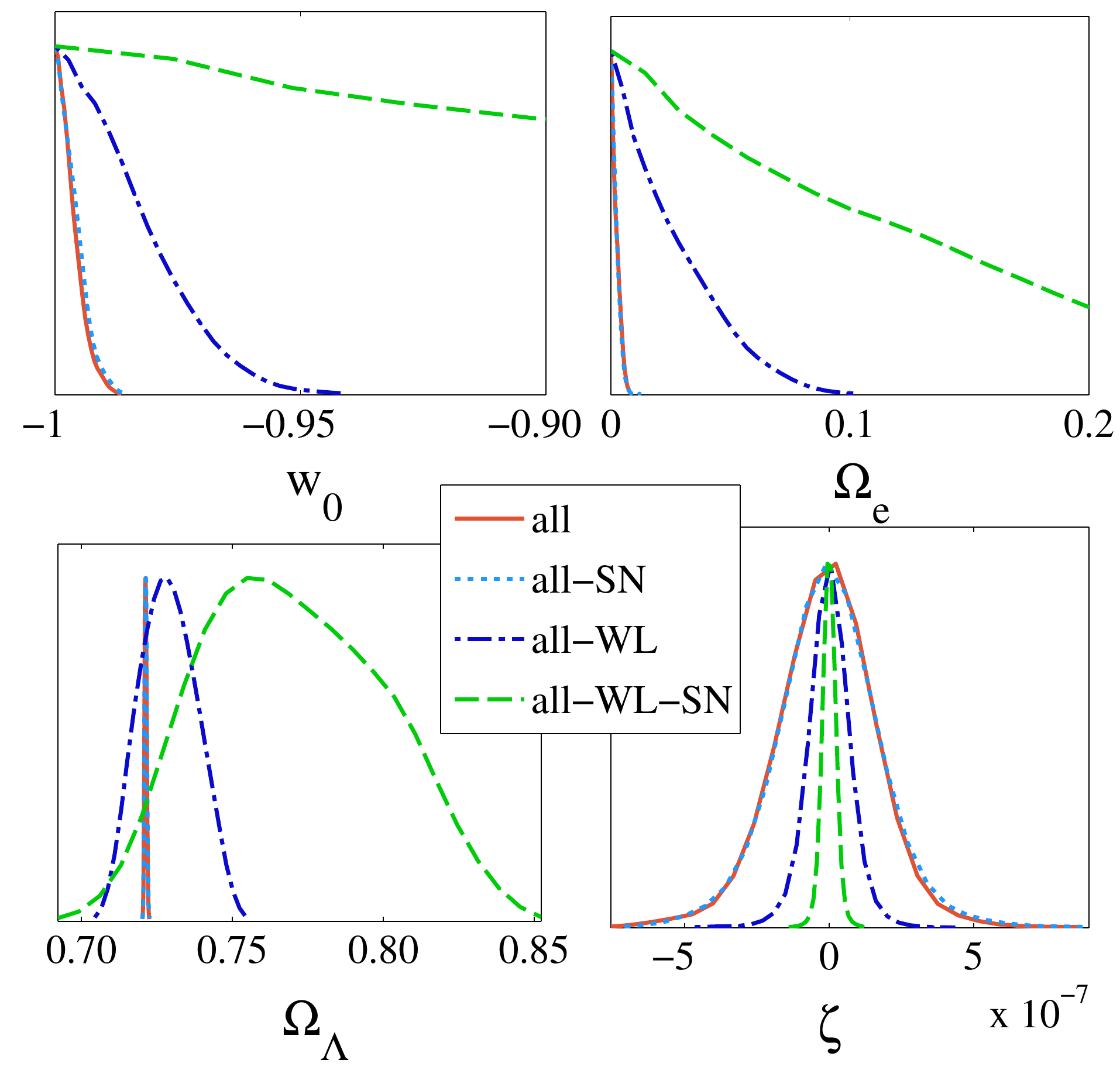}
\caption{Same as Fig.~\ref{fig:zetaCPL} for EDE parameters. }
\label{fig:zetaEDE}
\end{flushright}
\end{figure}

In the EDE case the considered low redshift combination of probes leads to very tight constraints on the model parameters, narrowing the parameter space in a competitive way with respect to present high redshift results on this kind of models (see \cite{act},\cite{pettorino} for latest results). We obtain $w_0 < -0.992$ and $\Omega_{\rm e} < 0.0051$ at 95\% c.l. and we report the 2-dimensional distribution in Fig.~\ref{fig:w0oe}. Moreover we can see in Fig.~\ref{fig:zetaEDE} that the effect on the coupling constraints discussed above for CPL holds also when the $\alpha$ variation is driven by this kind of dark energy parametrization: the more datasets we consider, the broader the constraints on the coupling are.

\begin{figure*}[htb!]
\begin{flushright}
\begin{tabular}{cc}
\includegraphics[width=9cm]{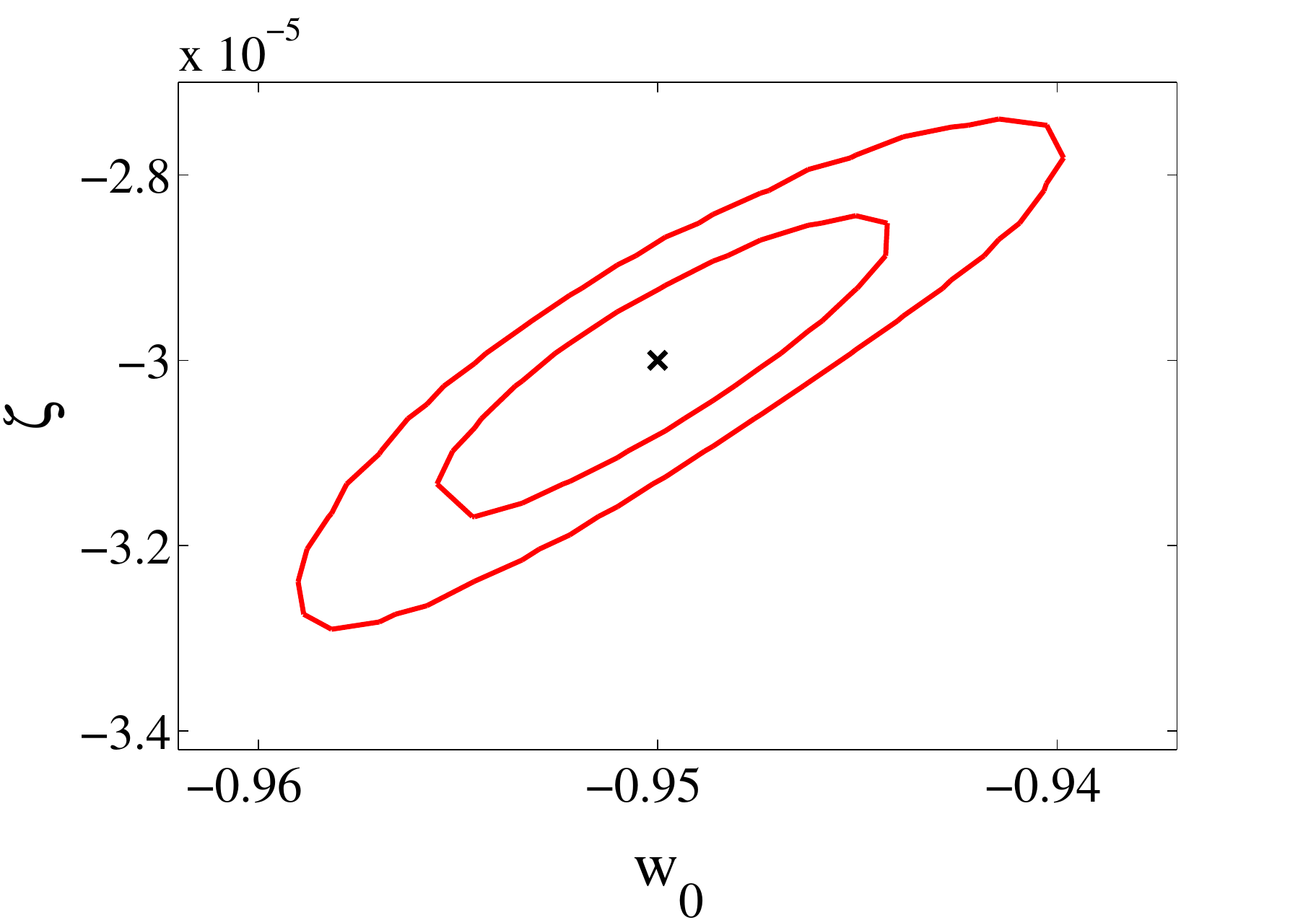} &
\includegraphics[width=9cm]{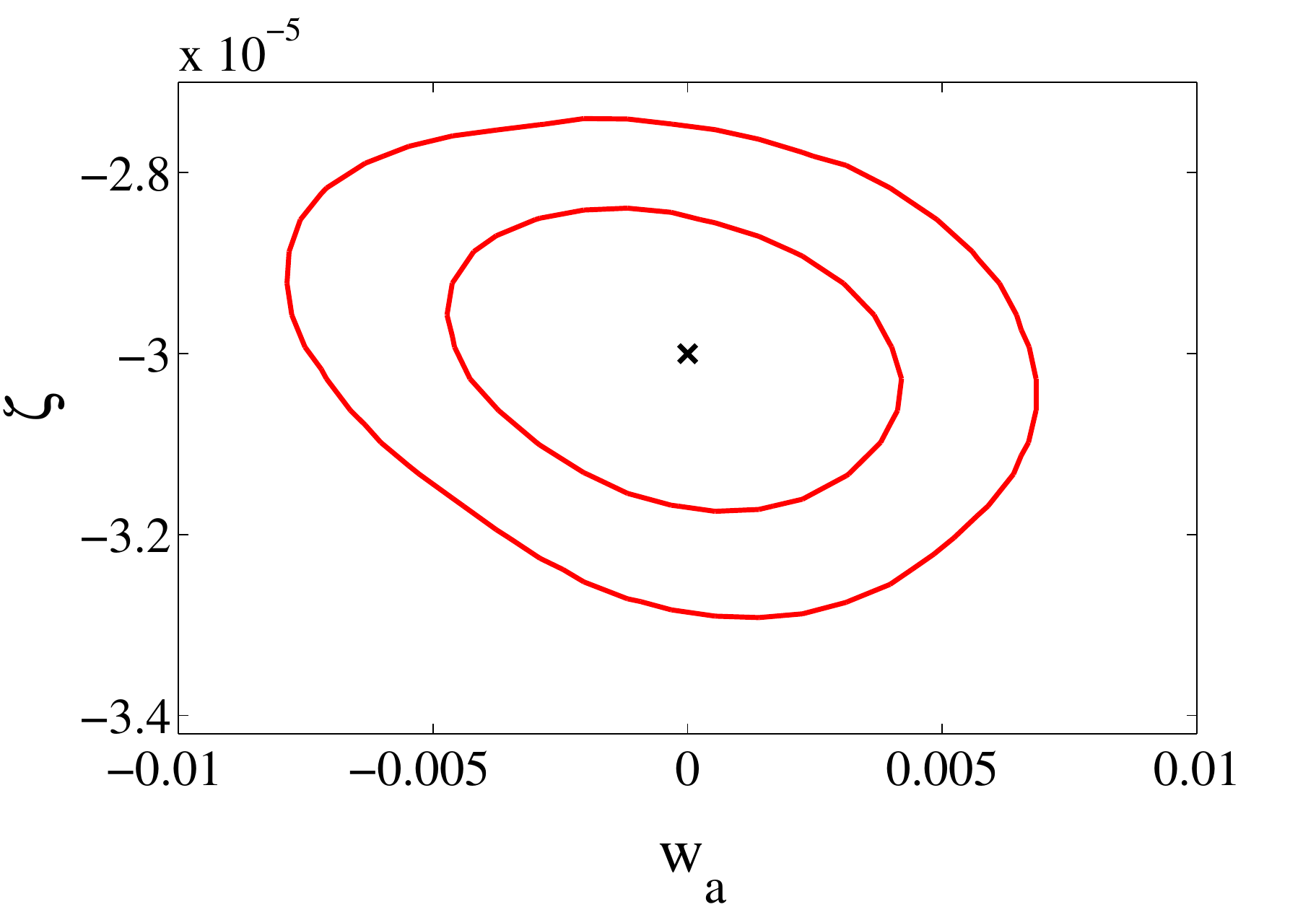}\\
\includegraphics[width=9cm]{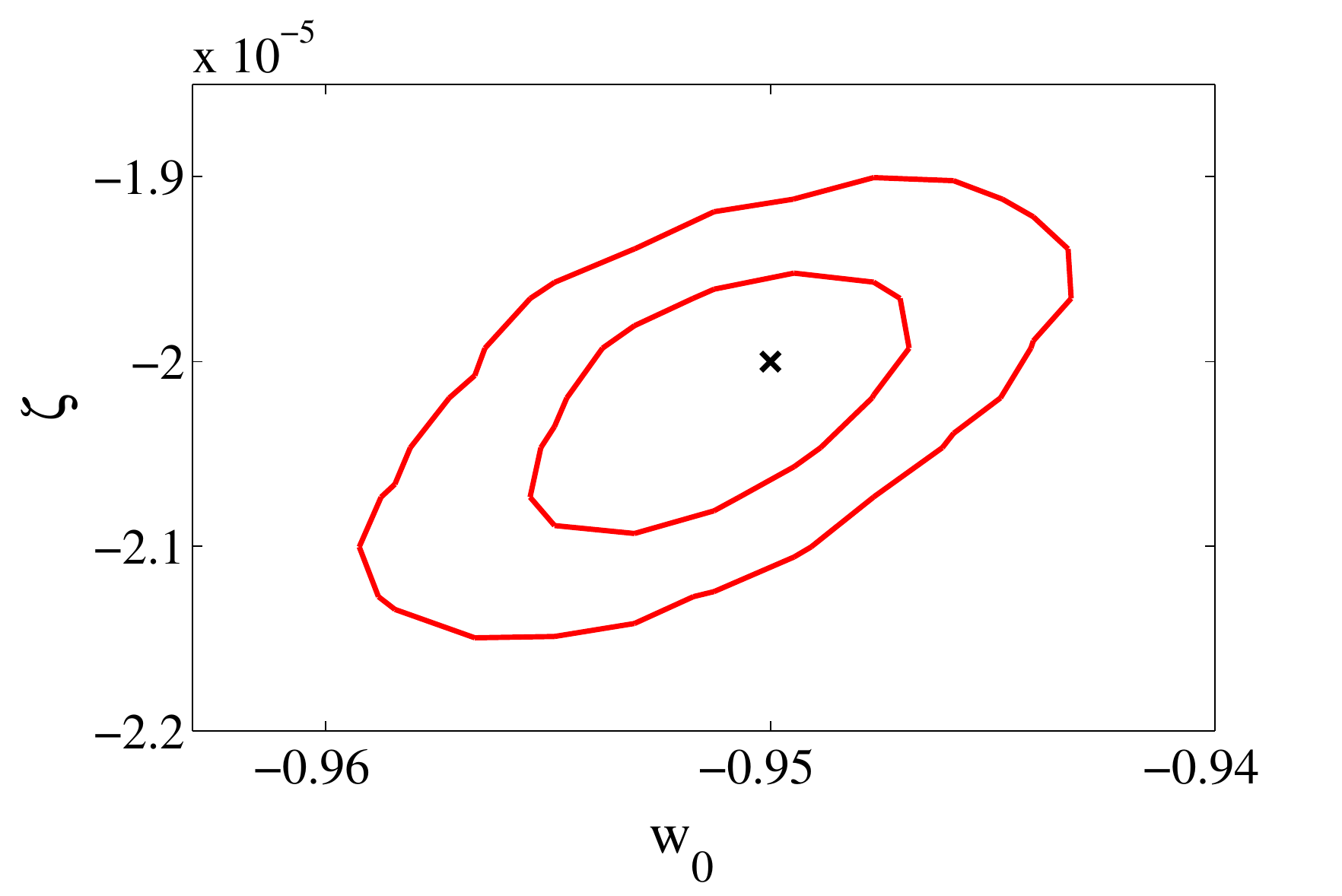} &
\includegraphics[width=9cm]{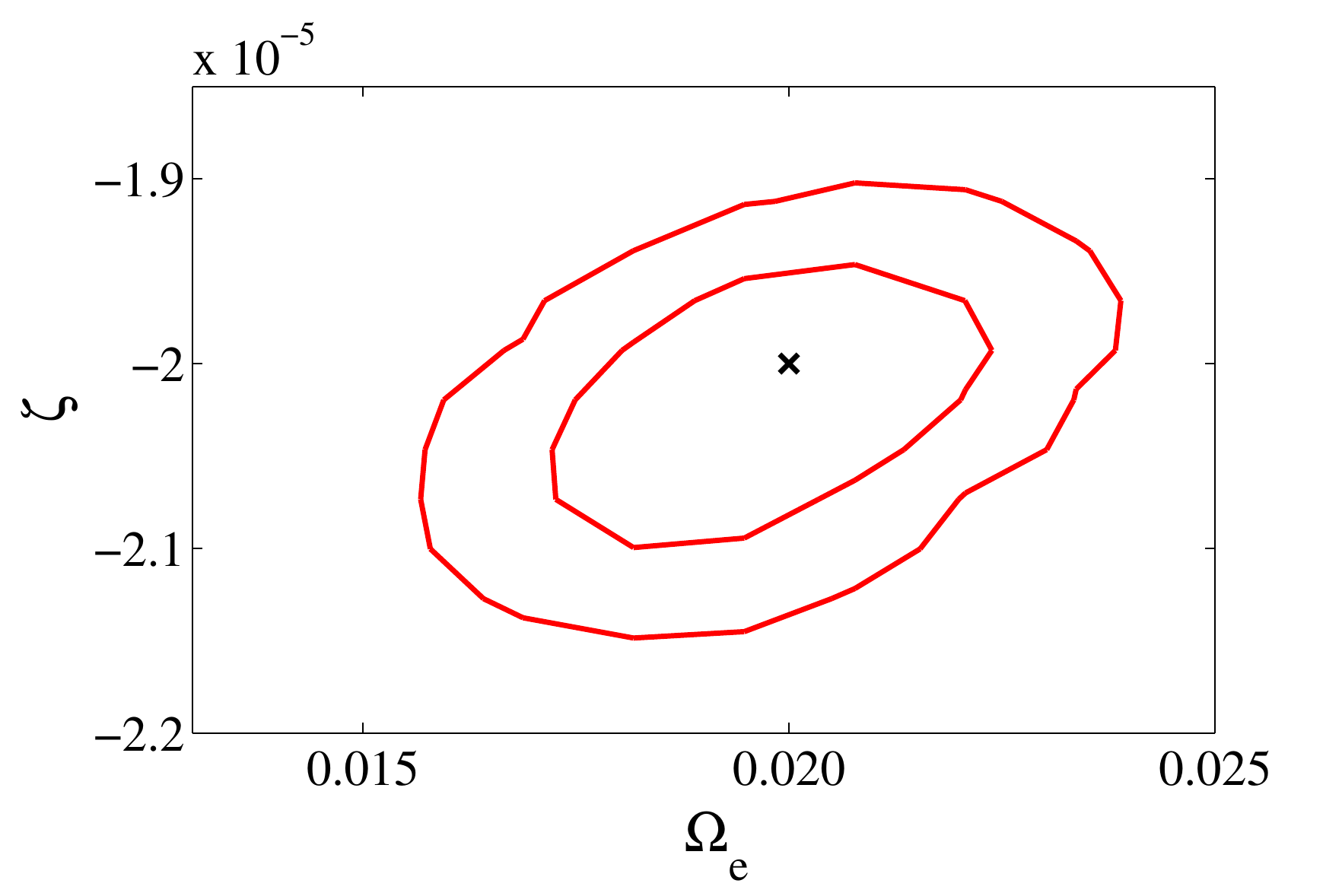} \\
\end{tabular}
\caption{{\it Top panels}: 2-dimensional contours at 68\% and 95\% confidence levels showing $\zeta$ versus $w_0$/$w_a$ for the CPL model when a Set2 fiducial cosmology is assumed in the data building. {\it Bottom panels}: same as top panel showing $\zeta$ versus $w_0$/$\Omega_e$ EDE model parameters. The black crosses show the chosen fiducial values. }
\label{fig:DEdeg}
\end{flushright}
\end{figure*}
\subsection{Non-Vanishing $\Delta\alpha/\alpha$}
In a second step of our analysis we select fiducial models (Set2, Set3, and Set4) where \dalfa is not vanishing and the DE parameters move from the standard $\Lambda$CDM scenario. We report constraints on DE parameters for both the CPL and EDE parametrizations, as well as for the coupling arising in a BSBM model.

In this case, the peculiar $w-\zeta$ behaviour mentioned above, due to the $\zeta=0$ fiducial value, is not present and the degeneracies between these parameters show up clearly, as we report in Fig.~\ref{fig:DEdeg} for both CPL and EDE models. 

\begin{figure}[htb!]
\begin{center}
\includegraphics[width=\columnwidth]{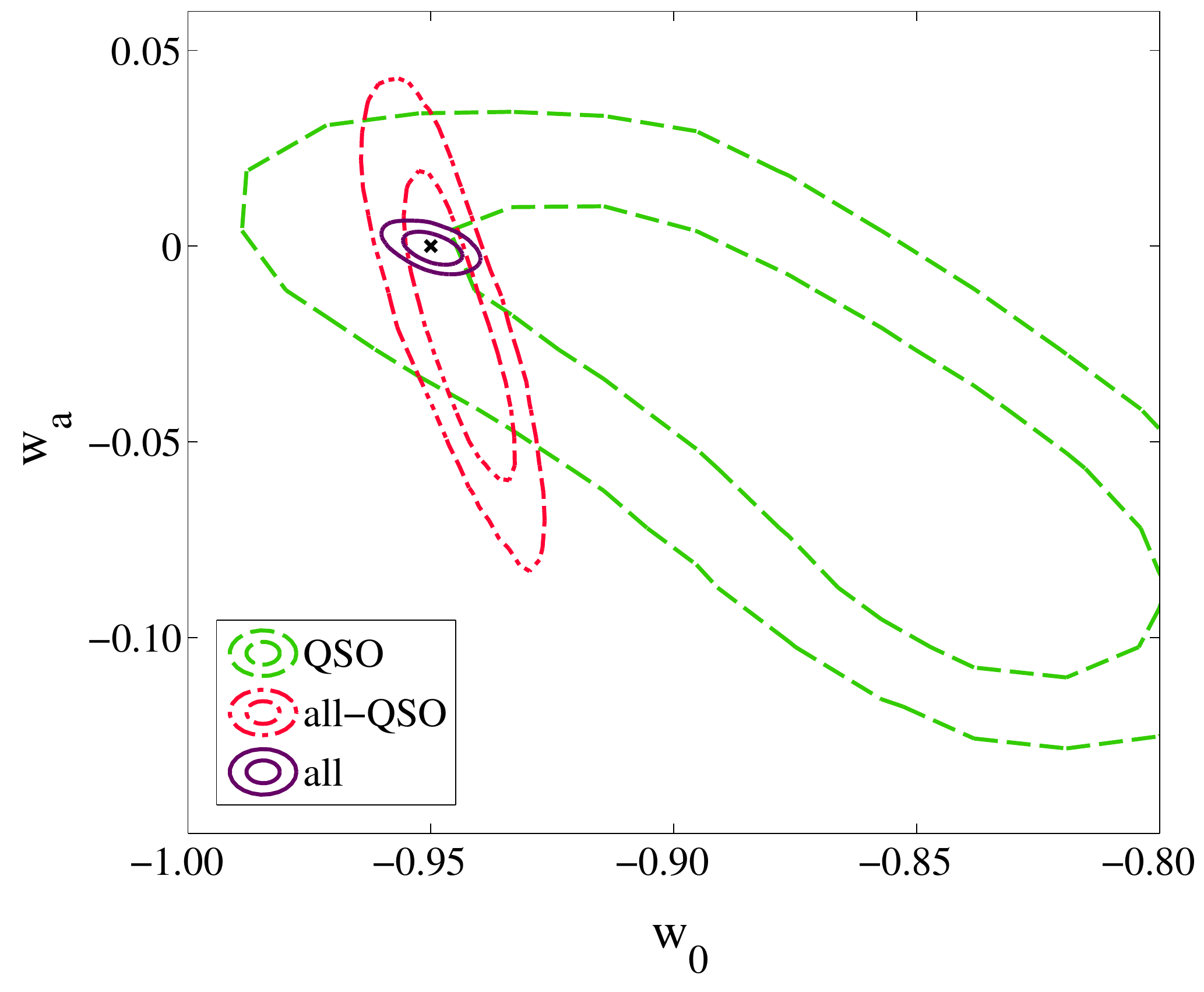}\\
\caption{QSO contribution to the $w_0$-$w_a$ constraints. We report contour plots at 68\% and 95 \% confidence levels as obtained from QSO data only (dashed green line), all probes except QSO (dash-dotted red line) and all probes (solid purple line). The black cross shows the fiducial input values.}
\label{fig:qso}
\end{center}
\end{figure}

We also notice that probing a different fiducial cosmology will give different constraints on the parameters. For the CPL parametrization we recover the input fiducial values and we obtain $\sigma(w_0)=0.004$, $\sigma(w_a)=0.003$ and $\sigma(\zeta)=1.1\times 10^{-6}$. The constraint on $w_0$ improves by a factor of about two and the measurement of $w_a$ becomes about one order of magnitude better: moving the fiducial region away from the special point ($\zeta=0$, $w_0=-1$) prevents the loss of constraining power because of the pathological degeneracies described in Fig.~\ref{fig:trendCPL} and therefore all the observables can fully contribute in constraining the cosmological parameters. In particular, in these non standard scenarios, the QSO contribution will be non vanishing.  Even though QSO data have a much lower constraining power than other dark energy observables, in Fig.~\ref{fig:qso} it is possible to notice how this dataset can provide independent (and almost orthogonal) limits on dark energy parameters and can be used to break degeneracies between $w_0$ and $w_a$. 

The same behaviour is observed in the EDE analysis where we find $\sigma(w_0)=0.003$, $\sigma(\Omega_e)=0.001$ and $\sigma(\zeta)=5.0\times 10^{-7}$ at 68\% c.l.; the EDE parameters will be detected with high significance in this scenario.

\begin{figure}[ht!]
\begin{center}
\includegraphics[width=\columnwidth]{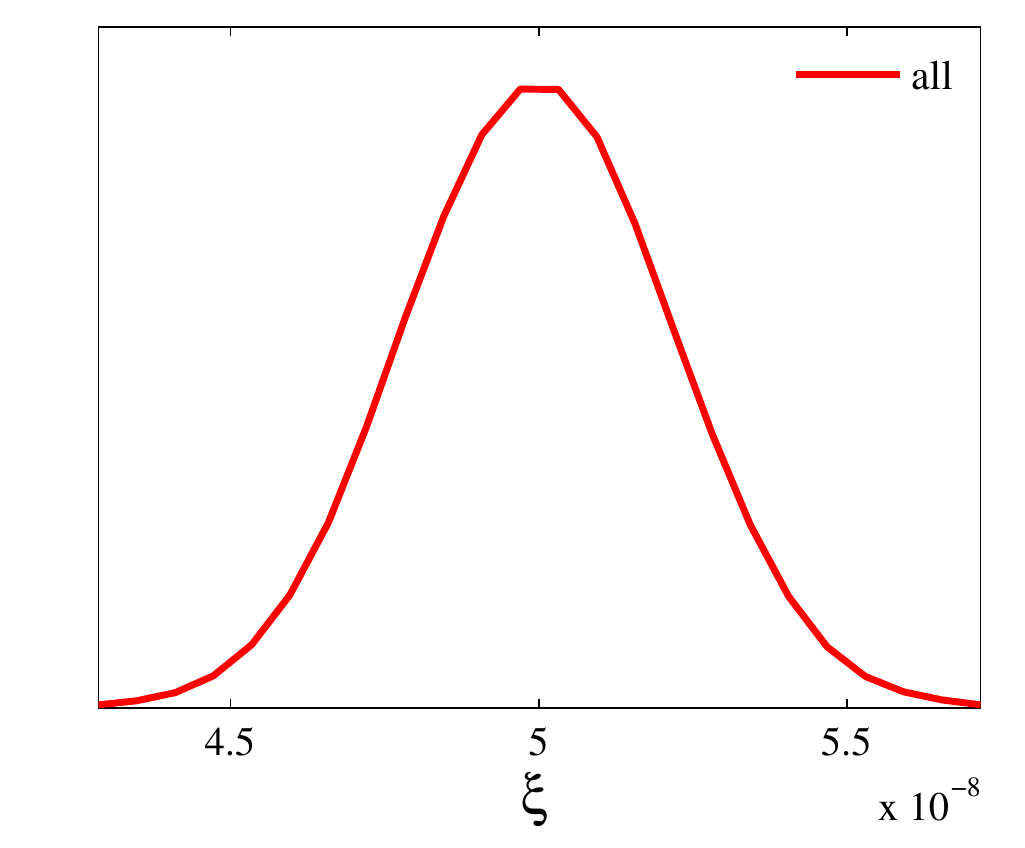}
\caption{Marginalized 1-dimensional posterior distribution for the coupling parameter $\xi$ between the BSBM scalar field and $\alpha$. This result refers to the combination of all the considered datasets.}
\label{fig:bsbm}
\end{center}
\end{figure}

Set4 defines the non vanishing \dalfa fiducial model used to forecast the coupling between the electromagnetic sector and the BSBM scalar field which, as explained above, does not affect the background expansion of the universe. This implies that probes which do not directly depend on $\alpha$ will constrain cosmological parameters but will not be sensitive to the coupling $\xi$ in any case, given that Eq.(\ref{eq:bsbm}) relies only on $\xi$ as free parameter. Therefore in this analysis $\xi$ is constrained only by QSO and SN data, the latter through the shift a variation of $\alpha$ produces on the distance modulus. We constrain $\sigma(\xi)=2.1\times 10^{-9}$ (see Fig.~\ref{fig:bsbm}).

As a last investigation we analyse the non vanishing $\alpha$ data fixing the coupling parameter to zero in the cosmological parameter estimation. This assumption will force the analysis to fit datasets where $\Delta\alpha/\alpha$ is redshift dependent with theoretical spectra unable to reproduce this trend. Should this translate into a bias in the recovered cosmological parameters we will be able to quantify the impact of a wrong assumption on $\zeta$ on cosmological results. 

\begin{figure}[htb!]
\begin{center}
\includegraphics[width=\columnwidth]{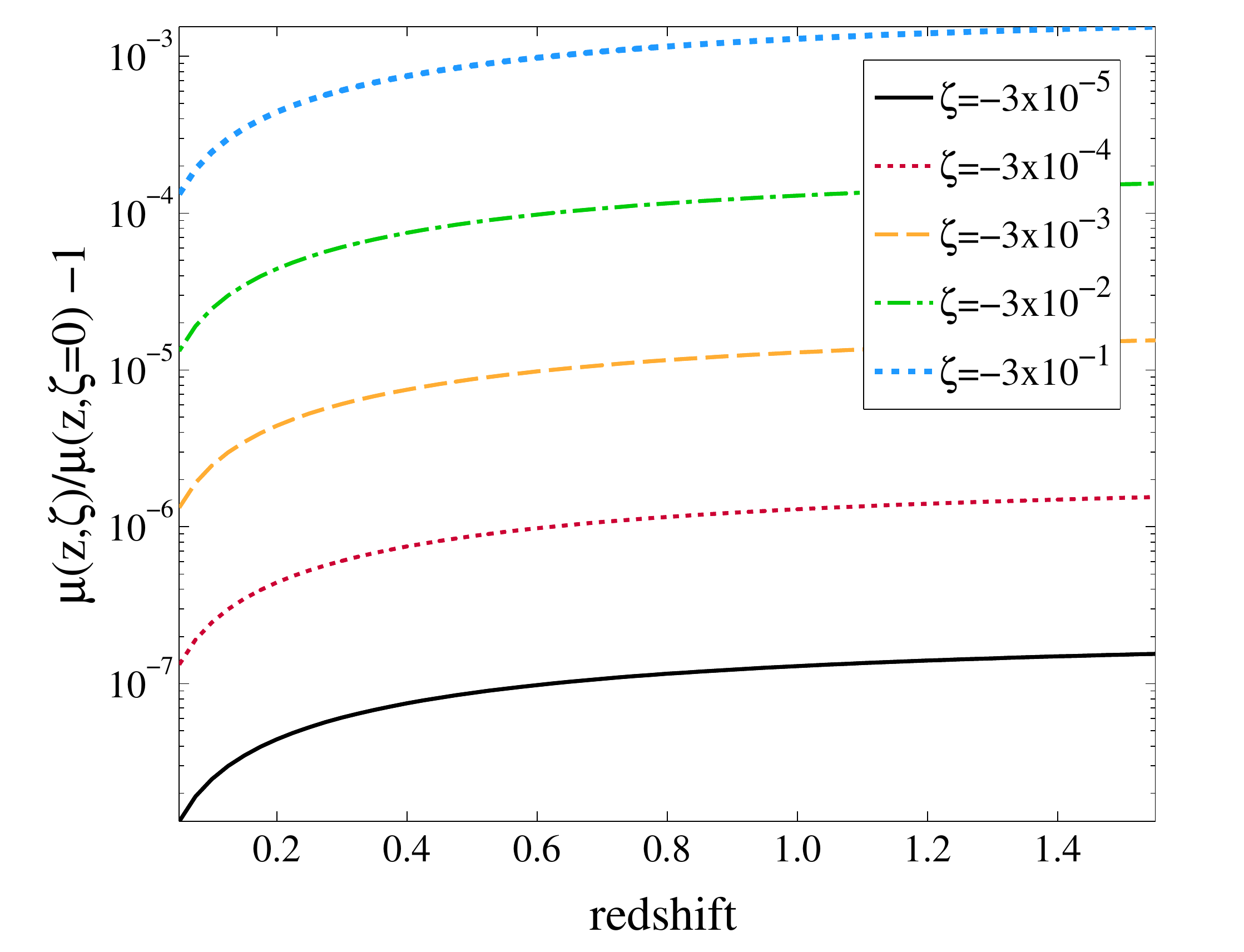}\\
\caption{Distance modulus $\mu(z)$ produced in CPL cosmology for different values of  $\zeta$, compared to the fiducial case with $\zeta=0$. We see that the greater is the departure from $\zeta=0$, the greater the shift in $\mu(z)$ will be.}
\label{fig:shiftSN}
\end{center}
\end{figure}

Among the observables we considered in this work, only QSO and SN are directly affected by the $\alpha$ evolution, and in particular only SN can produce a shift in the estimated value of the cosmological parameters. $\zeta=0$ will in fact always produce a vanishing \dalfa in  Eq.(\ref{eq:dalfa}). Thus, whatever value the cosmological parameters assume, the whole parameter sets will not give a good fit to the QSO dataset which directly probe the quantity \dalfa. On the contrary, SN datasets generated with $\Delta\alpha/\alpha\neq0$ are shifted with respect to the Set1 dataset (see Eq.(\ref{eq:snalfa})), and require a shift in the cosmology affecting $\mu_0(z)$ to compensate this artefact. We better show this effect in Fig.\ref{fig:shiftSN} where we plot the relative difference of the distance modulus $\mu(z)$ for different coupling values with respect to the case $\zeta=0$ for the CPL model. We see that the greater is the departure from $\zeta=0$, the greater the shift in $\mu(z)$ will be. 

Nevertheless we find that, assuming Set2, Set3 and Set4 fiducial values, this bias is too small to be observed with the considered SN survey in both types of models. We do not find any significant shift in the cosmological parameters induced by wrong assumptions on the coupling, suggesting that a more sensitive and deep SN survey will be needed to detect this effect. Indeed the E-ELT (plus JWST \cite{Whalen:2012yk}) is expected to find SN up to $z\sim5$ probing the region at higher $z$ where the shift in $\mu(z)$ is slightly increasing. A greater value of $\zeta$ might have an effect as well however, as stated previously, we restrict our analyses to a parameter region in agreement with current observations, i. e. $|\zeta \lesssim 10^{-3}|$. 

\section{CONCLUSIONS}
\label{sec:vi}
In this paper we focused on the possible coupling between a scalar field driven dark energy, parametrized here with the CPL and EDE formalisms, and electromagnetism, which can in principle bring to a time evolution of the fine structure constant $\alpha$. We have shown how the two sectors are connected by a coupling $\zeta$ and we investigated the ability of future low-medium redshift surveys to constrain this coupling. In particular, we considered two different scenarios, a standard $\Lambda$CDM one (without $\alpha$ variations) and dynamical dark energy where a $\zeta\neq0$ produces a redshift evolution for the fine structure constant. We forecasted observables for these two fiducial cosmologies from several upcoming surveys and we analysed these simulated datasets using MCMC techniques.

In the vanishing \dalfa case we obtained constraints on the sampled parameters, showing how, as expected, dark energy parameters will greatly benefit from weak lensing and supernova data coming from the Euclid satellite: we find ($\sigma(w_0) = 0.007$, $\sigma(w_a)=0.03$) at 68\% c.l. and  ($w_0 < -0.992$, $\Omega_{\rm e} < 0.0051$) at 95\% c.l., for the CPL and EDE models respectively. Alongside this expected result, we also observe a rather peculiar behaviour on $\zeta$: the chosen fiducial cosmology in fact implies that the better dark energy parameters are constrained, the larger the range of allowed values for $\zeta$ is. When all observables are considered we get $\sigma(\zeta_{\rm CPL})=1.8\times 10^{-7}$ and $\sigma(\zeta_{\rm EDE})=1.7\times 10^{-7}$ at 68\% c.l..

This trend disappears when the second fiducial model is considered, as we move away from the peculiar point $[\zeta,w(z)]=[0,-1]$ of the parameter space. 
In the non-$\Lambda$CDM fiducial cosmology, we have shown the constraining power of the considered observables on the sampled parameters, as well as the degeneracies between dark energy parameters and $\zeta$ both for the CPL and EDE models, highlighting how these degeneracies affect constraints. In particular we showed for the CPL model how the contribution from QSOs, combined with orthogonal constraints from Euclid observables, will improve the estimate by a factor of $2$ for $w_0$ and by one order of magnitude for $w_a$, finding ($\sigma(w_0)=0.004$, $\sigma(w_a)=0.003$) at 68\% c.l.. A detection of dark energy parameters at high significance is predicted also in the EDE model, with ($\sigma(w_0)=0.003$, $\sigma(\Omega_e)=0.001$) at 68\% c.l.. The coupling is constrained with 
$\sigma(\zeta_{\rm CPL})=1.1\times 10^{-6}$ and $\sigma(\zeta_{\rm EDE})=5.0\times 10^{-7}$ at 68\% c.l..

Furthermore, we analysed this last fiducial cosmology keeping $\zeta$ fixed to a value different from the one in input in order to find out if wrong assumptions on the cosmological model could produce an observable bias on parameters. We discovered this is not the case as only SN can highlight this shift and the survey considered here is not sensitive enough to show this small effect. A future paper may investigate which are the specifications (such as the number of SN and redshift range) needed by a future survey to detect this bias.

Finally, we also considered a BSBM model, where the scalar field coupled to electromagnetism is not the one driving the accelerated expansion of the universe. We analysed this model using datasets forecasted with a fiducial cosmology producing a non vanishing \dalfa and obtained constraints on the coupling of this model with electromagnetism, obtaining $\sigma(\xi)=2.1\times10^{-9}$. Also in this case we investigated the possible existence of a bias due to wrong cosmological assumptions, finding the same results obtained for the CPL and EDE models.\\

\section*{AKNOWLEDGMENTS}
We would like to thank Silvia Galli for useful discussions. We acknowledge useful comments and suggestions from Isobel Hook and other members of the Euclid Cosmology Theory SWG. EC acknowledges funding from ERC grant 259505. MM acknowledges partial support from the PD51 INFN grant. CJM and PEV acknowledge funding from the project PTDC/FIS/111725/2009 (FCT, Portugal). CJM is also supported by an FCT Research Professorship, contract reference IF/00064/2012, funded by FCT/MCTES (Portugal) and POPH/FSE (EC). The Dark Cosmology Centre is funded by the Danish National Research Foundation. SS acknowledges the support of ASI contract n. I/023/12/0. VFC is funded by Italian Space Agency (ASI) through contract Euclid-IC (I/031/10/0) and acknowledges financial contribution from the
agreement ASI/INAF/I/023/12/0.

\appendix
 \begin{table*}[htb!]
 \centering
 \begin{tabular}{c|c|c|c|c|c}
 \hline
& \multicolumn{5}{c}{CPL} \\
\hline
 & all & all-WL & all-SN & all-RD & all-QSOCL \\
 \hline
 $\sigma(\Omega_bh^2)$    &   $ 5.4\times10^{-4}$  & $< 0.025$ & $<0.025$ & $5.3\times10^{-4}$ & $5.4\times10^{-4} $\\
 $\sigma(\Omega_ch^2)$    &   $6.6\times10^{-4} $  & $4.8\times10^{-3}$ & $3.3\times10^{-3}$& $6.6\times10^{-4}$ & $6.9\times10^{-4}$\\
 $\sigma(H_0)$                  &   $1.6\times10^{-2} $  & $2.1 \times10^{-2}$ & $1.2$& $1.6\times10^{-2}$ & $1.6\times10^{-2}$\\
 $\sigma(\Omega_\Lambda)$    &   $6.8\times10^{-4} $ & $9.3 \times10^{-3}$ & $8.9\times10^{-4}$& $6.9\times10^{-4}$& $7.8\times10^{-4}$\\
 $\sigma(w_0)$                   &   $6.8\times10^{-3}$  & $1.6\times10^{-2}$ & $9.2\times10^{-3}$& $6.8\times10^{-3}$& $7.5\times10^{-3}$\\
 $\sigma(w_a)$                   &   $2.6\times10^{-2}$   & $9.8\times10^{-2}$ & $3.5\times10^{-2}$& $2.6\times10^{-2}$& $3.1\times10^{-2}$\\
 $\sigma(\zeta)$             &   $1.8\times10^{-7} $ & $9.7 \times10^{-8}$ & $1.6\times10^{-7}$& $1.8\times10^{-7}$& $1.0<\times10^{-5}$\\
 \hline
 \end{tabular}
 \vspace{1cm}

 \begin{tabular}{c|c|c|c|c|c}
 \hline
&\multicolumn{5}{c}{EDE} \\
\hline
 & all & all-WL & all-SN & all-RD & all-QSOCL\\
 \hline
$\sigma(\Omega_bh^2)$    &  $5.4\times10^{-4}$ & $<0.025$ & $<0.025$& $5.3\times10^{-4}$ & $5.3\times10^{-4}$\\
$\sigma(\Omega_ch^2)$    &   $6.2\times10^{-4}$  & $<0.12$ & $<0.12$& $6.2\times10^{-4}$ & $6.1\times10^{-4}$\\
$\sigma(H_0)$                  &   $1.3\times10^{-2}$  & $2.2\times10^{-2}$ & $1.2$& $1.3\times10^{-2}$ & $1.3\times10^{-2}$\\
$\sigma(\Omega_\Lambda)$    &   $3.5\times10^{-4}$ & $9.2\times10^{-3}$ & $3.7\times10^{-4}$& $3.6\times10^{-4}$ & $3.7\times10^{-4}$\\
$\sigma(w_0)$                   &   $<-0.996$  & $<-0.983$ & $<-0.996$& $<-0.996$ & $<-0.996$\\
$ \sigma(\Omega_e)$         &  $<2.6\times10^{-3}$& $<2.9\times10^{-2}$ & $<2.4\times10^{-3}$& $<2.6\times10^{-3}$ & $<2.8\times10^{-3}$\\
$\sigma(\zeta)$             &   $1.7\times10^{-7}$ & $7.8\times10^{-8}$ & $1.7\times10^{-7}$& $1.9\times10^{-7}$ & $<1.0\times10^{-5}$\\
 \hline
 \end{tabular}
 \caption{$68 \%$ c.l. constraints on relevant cosmological parameters when the DE equation of state is parametrized through the CPL (top) or EDE (bottom) formalism for different combinations of probes: \emph{all} includes all datasets described in Section \ref{sec:iii}, \emph{all-WL} excludes weak lensing data, \emph{all-SN} excludes supernovae data, \emph{all-RD} excludes redshift drift, and \emph{all-QSOCL} excludes quasars and atomic clocks bounds.}
 \quad
  \label{tab:resCPLEDE}
 \end{table*}

\section{Recovered Parameters}
\label{sec:app}
 While in Section~\ref{sec:v} we focused only on the key results of our analyses, in this appendix we list in Section~\ref{sec:van_all} the constraints on all the parameters as determined from different combinations of probes, and in Section~\ref{sec:nonvan} we report the 1-dimensional posteriors and the constraints for the sampled parameters when a non-standard fiducial model is assumed.
 \subsection{Vanishing \dalfa}\label{sec:van_all}
\begin{figure*}[htb!]
 \begin{center}
  \includegraphics[scale=0.41]{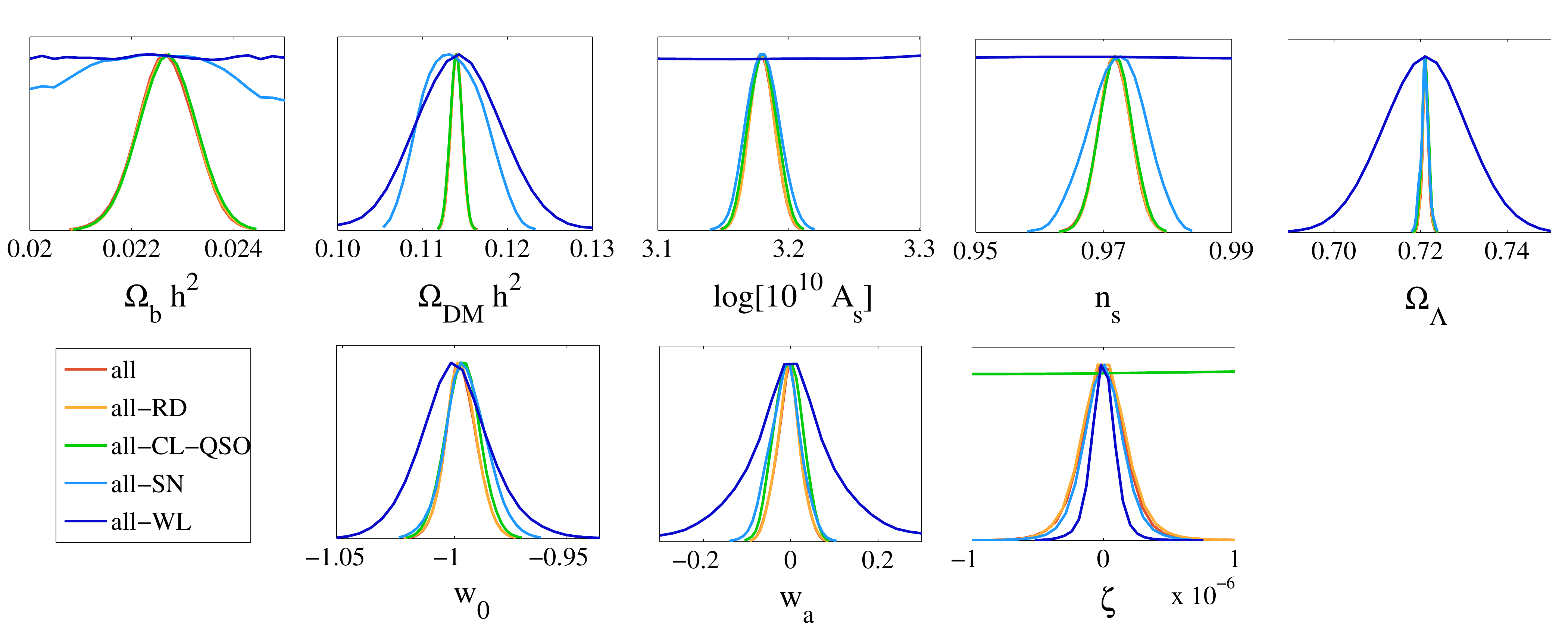}\\
 \includegraphics[scale=0.42]{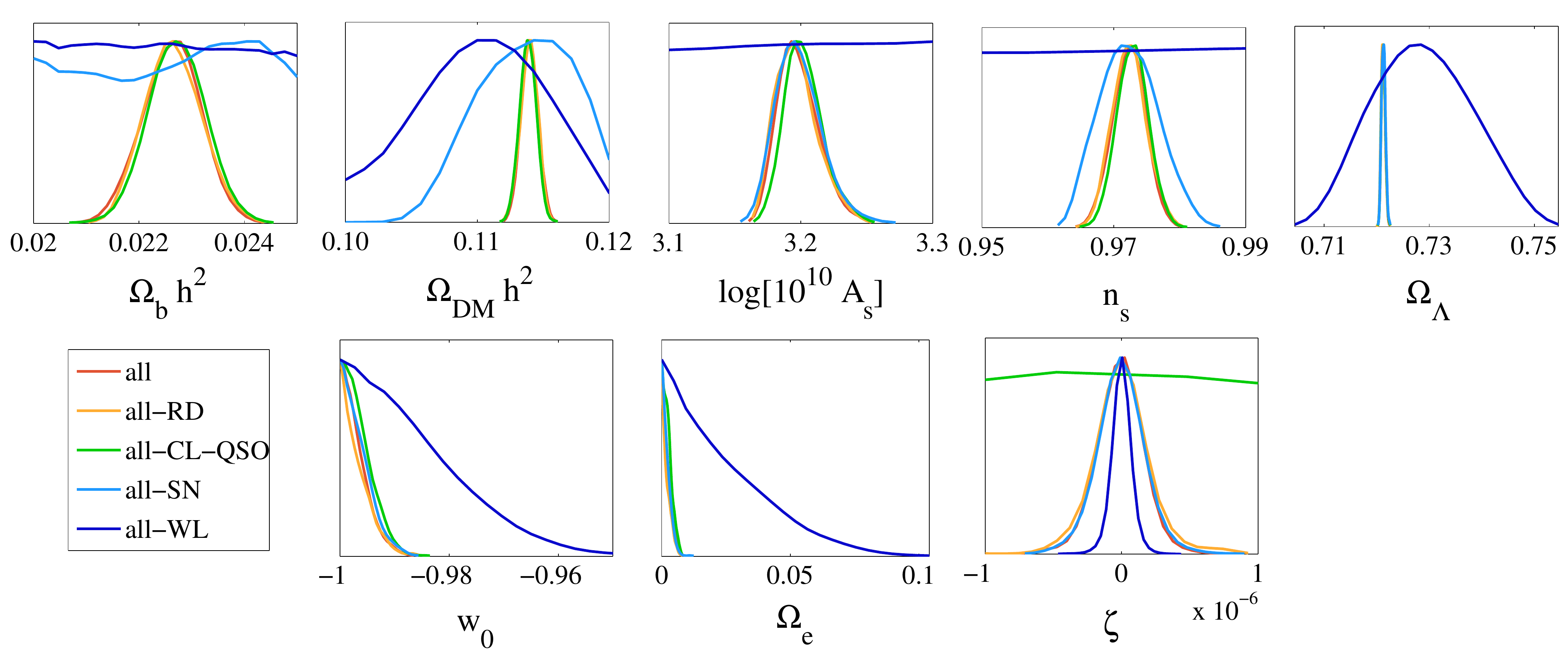}
 \caption{Marginalized posteriors for the sampled parameters in the analyses reported in Table~\ref{tab:resCPLEDE}. The top panel shows the results for a CPL parametrized DE, while the bottom panel refers to EDE results. Different curves in each panel refer to different combinations of probes : \emph{all} (red) includes all datasets described in Section \ref{sec:iii}, \emph{all-WL} (dark blue) excludes weak lensing data, \emph{all-SN} (light blue) excludes supernovae data, \emph{all-RD} (orange) excludes redshift drift, and \emph{all-CL-QSO} (green) excludes quasars and atomic clocks bounds. }
 \label{fig:resCPLEDE}
 \end{center}
 \end{figure*}

For vanishing \dalfa datasets we performed several analyses, excluding each time one of the observables presented in Section~\ref{sec:iii}. In this way we could explore and highlight the contribution of each observable to the constraints. In Table~\ref{tab:resCPLEDE} we report the 68\% confidence level errors on relevant cosmological parameters and in Fig.~\ref{fig:resCPLEDE} we show the 1-dimensional posteriors recovered for both CPL and EDE models. We can notice how removing QSO and atomic clocks from the analysis we lose, trivially, all the constraining power for the coupling $\zeta$, while as expected removing WL and/or SN opens the DE parameters.

 \subsection{Non vanishing \dalfa}\label{sec:nonvan}
 
\begin{figure*}[htb!]
 \begin{center}
 \includegraphics[width=\textwidth,height=6cm]{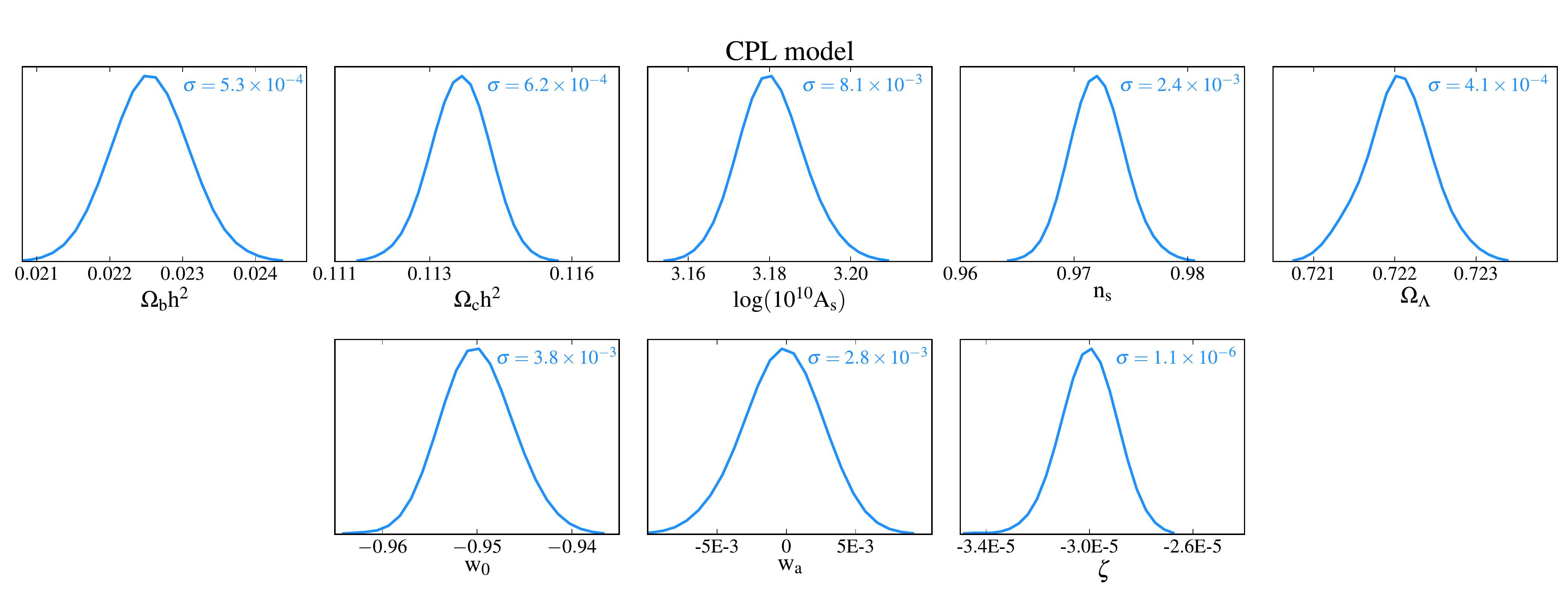}\\
 \includegraphics[width=\textwidth,height=6cm]{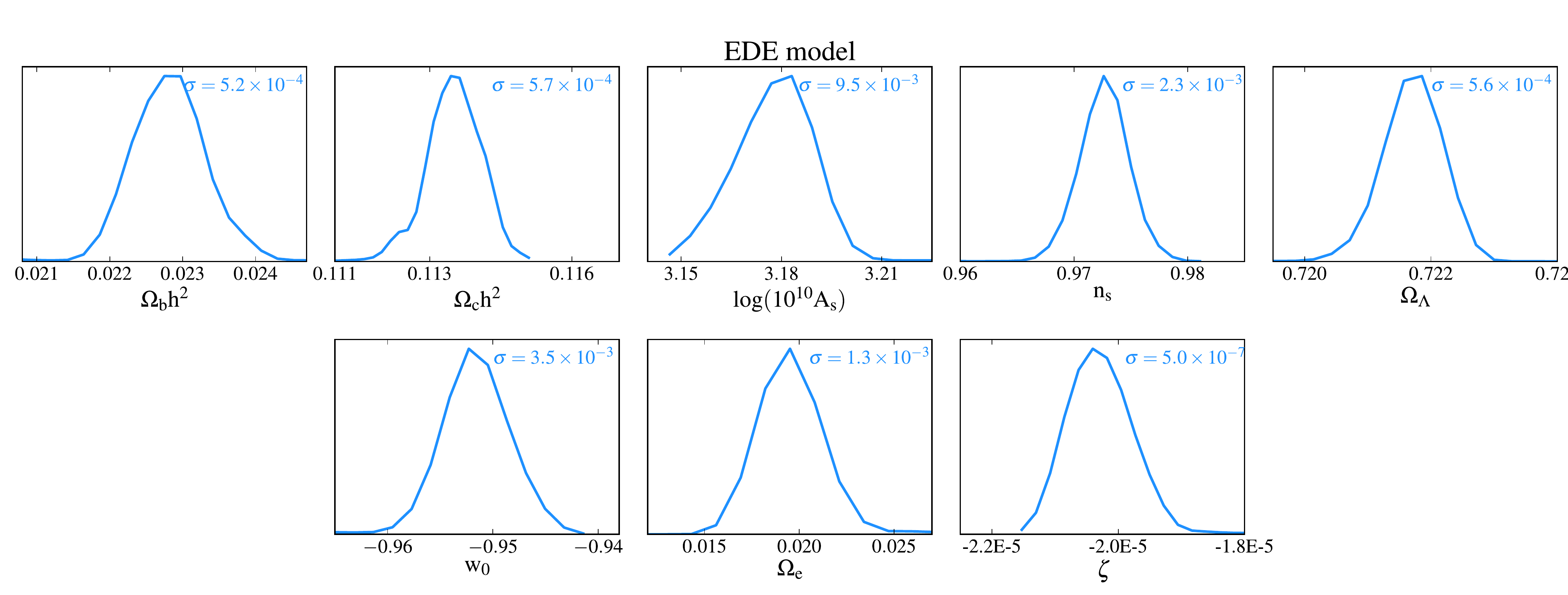}\\
\includegraphics[width=\textwidth,height=6cm]{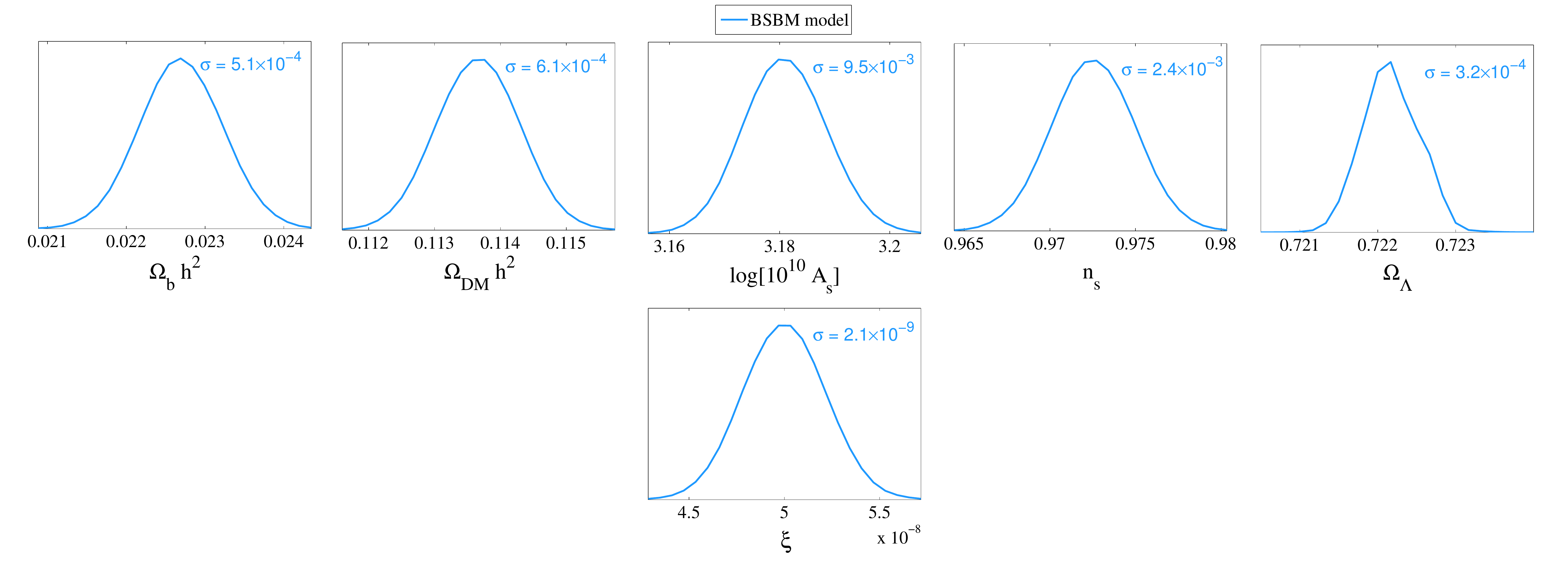}\\
 \caption{Marginalized posteriors for the sampled parameters when we assume Set2, Set3 and Set4 fiducial models (see Table~\ref{tab:fid_nost}). The top panel shows the results for a CPL parametrized DE, in the middle panel we report EDE results, and in last bottom panel we show the BSBM model. We also report the 68\% c.l. errors recovered on the parameters. }
 \label{fig:shifts}
 \end{center}
 \end{figure*}

In this section we report the constraints obtained when a non standard $\Lambda$CDM cosmology and a non vanishing \dalfa is assumed (Set2, Set3, Set4 in Table~\ref{tab:fid_nost}). We report results in Fig.~\ref{fig:shifts} only for the combination of all the observables.


\end{document}